\documentclass[fleqn,usenatbib]{mnras}

\usepackage[T1]{fontenc}
\usepackage{ae,aecompl}

\usepackage{eso-pic}

\usepackage{graphicx}	
\usepackage{amsmath}	
\usepackage{amssymb}	
\usepackage[shortlabels]{enumitem}

\usepackage{multicol}

\title[MaNGA Fundamental Plane]{The Stellar Mass Fundamental Plane:  The virial relation and a very thin plane for slow-rotators}

\author[ M.~Bernardi et al.]{
\parbox{\textwidth}{
  M.~Bernardi$^{1}$\thanks{\texttt{\rm \texttt{bernardm@sas.upenn.edu}}}, H.~Dom\'{i}nguez S{\'a}nchez$^{1,2}$,  B. Margalef-Bentabol$^{1}$, F. Nikakhtar$^{1}$ and R.~K.~Sheth$^{1}$}\\
 \vspace{0.cm}\\~\\
$^{1}$ Department of Physics and Astronomy, University of Pennsylvania, Philadelphia, PA 19104, USA\\
$^{2}$ Institute of Space Sciences (ICE, CSIC), Campus UAB, Carrer de Magrans, E-08193 Barcelona, Spain\\
}

\pubyear{2020}

\begin{document}
\label{firstpage}
\pagerange{\pageref{firstpage}--\pageref{lastpage}}
\maketitle

\begin{abstract}
  Early-type galaxies -- slow and fast rotating ellipticals (E-SRs and E-FRs) and S0s/lenticulars -- define a Fundamental Plane (FP) in the space of half-light radius $R_e$, enclosed surface brightness $I_e$ and velocity dispersion $\sigma_e$.  Since $I_e$ and $\sigma_e$ are distance-independent measurements, the thickness of the FP is often expressed in terms of the accuracy with which $I_e$ and $\sigma_e$ can be used to estimate sizes $R_e$. We show that:  1) The thickness of the FP depends strongly on morphology.  If the sample only includes E-SRs, then the observed scatter in $R_e$ is $\sim 16\%$, of which only $\sim 9\%$ is intrinsic. Removing galaxies with  $M_*<10^{11}M_\odot$ further reduces the observed scatter to $\sim 13\%$ ($\sim 4\%$ intrinsic).  The observed scatter increases to the $\sim 25\%$ usually quoted in the literature if E-FRs and S0s are added.  If the FP is defined using the eigenvectors of the covariance matrix of the observables, then the E-SRs again define an exceptionally thin FP, with intrinsic scatter of only 5\% orthogonal to the plane.
  2) The structure within the FP is most easily understood as arising from the fact that $I_e$ and $\sigma_e$ are nearly independent, whereas the $R_e-I_e$ and $R_e-\sigma_e$ correlations are nearly equal and opposite.
  3) If the coefficients of the FP differ from those associated with the virial theorem the plane is said to be `tilted'.  If we multiply $I_e$ by the global stellar mass-to-light ratio $M_*/L$ and we account for non-homology across the population by using S\'ersic photometry, then the resulting stellar mass FP is less tilted.  Accounting self-consistently for $M_*/L$ gradients will change the tilt. The tilt we currently see suggests that the efficiency of turning baryons into stars increases and/or the dark matter fraction decreases as stellar surface brightness increases.
\end{abstract}




\section{Introduction}\label{sec:intro}
Early-type galaxies (ellipticals and lenticulars) populate a manifold of lower dimensionality than that of the space of observables \citep{Brosche1973}.  They define a Fundamental Plane (FP) in the space of projected half-light radius $R_e$, enclosed surface brightness $I_e = (L/2)/(\pi R_e^2)$ and velocity dispersion $\sigma_e$ \citep{Djorgovski1987,comaFP,Jorgensen1996,sdssFP,SauronFP,SpiderII,6dFP}:
\begin{equation}
  R_e\propto \sigma_e^\alpha\,I_e^\beta,
  \label{eq:oldFP}
\end{equation}
with $\alpha \sim 1.2-1.4$ (depending on how the residuals to the plane are minimized -- these are often refered to as `direct' or `orthogonal' fits) and $\beta \sim -0.8$.  The Plane is usually expressed in this form -- rather than, e.g., $\sigma_e$ as a function of $R_e$ and $I_e$ -- because the quantities on the right hand side are distance-independent ($I_e$ scales as $(1+z)^4$, but the redshift $z$ is an observable).  

If galaxies are virialized systems, then a plane is not unexpected because the mass must scale as 
\begin{equation}
  M \propto k_n\frac{R_e\sigma_e^2}{G},
 \label{eq:Mdyn}
\end{equation}
where $k_n$ depends on the shape of the mass profile (the reason for the subscript $n$ will become clear shortly).  Using the fact that $L\propto I_eR_e^2$ and rearranging makes this 
\begin{equation}
 R_e\propto k_n\,\frac{\sigma_e^2}{I_{e}}\,\left(\frac{M_*}{L}\right)^{-1}
             \left(\frac{M_*}{M}\right),
 \label{eq:FP}
\end{equation}
where $M_*$ is the mass in stars.  Comparison with equation~(\ref{eq:oldFP}) shows that $\alpha=2$ and $\beta=-1$ are expected if $k_n$, the stellar mass to light ratio $(M_*/L)$, and the stellar-to-total mass fraction remain constant across the population.  Since the measured $\alpha$ and $\beta$ are very different from these values, at least one of these assumptions is incorrect.  The difference from the expected scaling is generally referred to as the `tilt' of the Fundamental Plane.  

Variations in $k_n$ indicate that galaxies are not homologous systems;
variations in $M_*/L$ imply changes in the star formation or assembly history; 
and variations in
\begin{equation}
  \frac{M_*}{M} = \frac{M_*}{M_{\rm bar}}\, \frac{M_{\rm bar}}{M_{\rm bar} + M_{\rm DM}}
  = \frac{M_*}{M_{\rm bar}}\,(1-f_{\rm DM})
 \label{eq:M*Mtot}
\end{equation}
imply changes in the overall efficiency of turning baryons into stars and in the dark matter fraction.  This has motivated studies of which of these variables contributes most to the tilt \citep{Bender92,Ciotti1996,Graves2009,SpiderII,6dFPsp,Cappellari2013a,Donofrio2017}.

As increasingly sophisticated estimates of $M_*/L$ became available, \cite{HB2009} showed that the coefficients $\alpha_*$ and $\beta_*$ in 
\begin{equation}
  R_e\propto \sigma_e^{\alpha_*}\,I_{*e}^{\beta_*}, \qquad{\rm where}\quad
  I_{*e} \equiv (M_*/L)\,I_e,
  \label{eq:HB*}
\end{equation}
are closer to the virial theorem values (they found $\alpha_* \approx 1.6$), leaving non-homology and variations in the dark matter fraction as reasons for the smaller remaining tilt.  One of our goals is to study what happens if we use recent improvements in S\'ersic-based photometry to at least partially account for the non-homology.\footnote{\cite{HB2009} used SDSS survey photometry because S\'ersic photometry for SDSS was not yet available and because dynamical mass estimates from SAURON \citep{SauronFP} suggested that galaxies were homologous, so this was not necessary.  However, recent work suggests that such dynamical mass estimates should be treated with caution \citep{Bernardi2018b,BDS2019}.}
I.e., we relate $k_n$ to the S\'ersic index $n$ \cite[following, e.g.][]{PS1997,Bernardi2018a}, and use this to study 
\begin{equation}
  \label{eq:FP*}
 \frac{R_e}{k_n}\propto \sigma_e^{\alpha_*}\,I_{*e}^{\beta_*}.
\end{equation}
In what follows we refer to the `traditional' Fundamental Plane defined by $I_e$, $R_e$ and $\sigma_e$ as FP$_{\rm L}$, while the Plane which results from replacing $I_e\to I_{*e}$ and $R_e\to R_e/k_n$ when S\'ersic parameters are used for $I_e$, $R_e$ and $k_n$ as the `stellar mass' Fundamental Plane FP$_{*}$. Note that FP$_*$ approximately isolates the contribution to the tilt which is due to $M_*/M$ of equation~(\ref{eq:M*Mtot}): If $\alpha_*\ne 2$ or $\beta_*\ne -1$, then this indicates that $M_*/M$ varies across the population.  

It is worth noting that $k_n$ in the discussion above assumes that $M_*/L$ gradients can be ignored.  If present, gradients will modify $k_n$ more than $I_{*e}$ \citep{Bernardi2018b}.  Figure~16 of \cite{BDS2019} suggests that $M_*/L$ gradients are present in MaNGA galaxies, but they are driven by IMF gradients.  Unfortunately, reliable estimates of IMF-driven effects require high S/N spectra, so evidence for gradients comes from stacked spectra.  As we do not have measured gradients for individual objects, we are unable to self-consistently correct $k_n$ and $I_{*e}$  here.\footnote{\cite{Chan2018} have recently used color gradients to approximate $M_*/L$ gradients.  We have not included such gradients in our analysis because it is the IMF driven gradients which matter most \citep{Bernardi2018b, BDS2019}, and these are not captured by color gradients.  In addition, while $M_*/L$ gradients based on analysis of the MaNGA spectra are available on an object-by-object basis from the MaNGA Firefly Value Added Catalog \citep{Goddard17}, these assume a fixed IMF so do not include IMF-driven gradients in $M_*/L$.}  Therefore, in addition to the effects mentioned above, some of the FP$_*$ tilt may arise from incorrectly ignoring gradients.

So far we have focussed on the tilt of the Fundamental Plane, as this is the quantity which encodes information about galaxy formation.  However, the thickness of the Plane is also interesting:  I.e. equations~(\ref{eq:oldFP}), (\ref{eq:HB*}) and~(\ref{eq:FP*}) show that $\sigma_e$, $I_e$ and $n$ can be used to predict $R_e$, but there will be some uncertainty in this prediction.  While the simple fact that the FP is tight constrains models of how early-type galaxies formed, the tightness is particularly interesting for cosmological studies.  This is because $\sigma_e$, $I_e$ and $n$ can be estimated for each galaxy without knowledge of its distance from us.  Since the angular half-light radius $\theta_e$ is observable, the rms difference between the predicted and true $R_e = \theta_e\,d_{\rm A}(z)$ is a measure of the uncertainty in the distances to the galaxies in the sample.  If the redshifts to the galaxies are known, this uncertainty in distance translates directly into an uncertainty in the velocity of the galaxies \cite[e.g.][]{efarVI,6dFPv}.  For more distant samples, gravitational lensing can affect the observed size $\theta_e$ -- but it does not change $I_e$ or $\sigma_e$ -- so the difference between the predicted size $\sigma^\alpha\,I_e^\beta$ and the true $R_e$ contributes to the uncertainty on an estimate of the lensing signal's effect on the observed size \citep{Bertin2006,Huff2014,Freudenburg2020}.  Peculiar velocities and gravitational lensing are powerful cosmological probes.  For this reason, there is considerable interest in samples for which the predicted size is as precise as possible.

One possible source of scatter is the morphological mix, as it is possible that, e.g., slow rotating ellipticals define a slightly different FP than, e.g., S0s which obviously have some rotational support.  E.g. the FP of Coma cluster galaxies appears to depend slightly on morphology \citep{comaFP}.  In what follows, we show how the FP depends on the morphological mix of the early-type galaxy sample.

Section~\ref{sec:data} describes the parent data set from which our sample is drawn, as well as the morphological classifications, and photometric and spectroscopic quantities we require for our analysis.  Section~\ref{sec:results} presents our results.
Section~\ref{sec:sdss} compares the MaNGA FP$_{\rm L}$ and FP$_*$s with those of Es in the SDSS.  A final section summarizes our findings.  Appendix~\ref{sec:sigVrot} describes a check of the importance of rotation on the estimated velocity dispersions we use when constructing the FP, and Appendix~\ref{sec:FPnorm} discusses the FP which results when all observables are first normalized by their rms values.

\section{Data: The MaNGA survey}\label{sec:data}
The MaNGA survey (\citealt{ Bundy2015}) is a component of the Sloan Digital Sky Survey IV (\citealt{Gunn2006, Smee2013, Drory2015, Blanton2017}; hereafter SDSS IV).  MaNGA uses integral field units (IFUs) to measure multiple spectra across $\sim$ 10000 nearby galaxies \cite[see][for the sample selection]{Wake2017}.
In this work, we use the MaNGA DR15 \citep{Aguado2019}, which provides kinematic maps (stellar rotation velocity and velocity dispersion) for each galaxy. All the objects in our sample are sufficiently distant that peculiar velocities make a negligble contribution to the measured redshifts.

\subsection{Photometry and morphology}
The photometric parameters and morphological classifications for the galaxies that we use below come from the MaNGA PyMorph Photometric Value Added Catalogue (MPP-VAC) and the MaNGA Deep Learning Morphological Value Added Catalogue (MDLM-VAC) presented in \citet{Fischer2019}. 

In this work, for each galaxy, we use the best-fit parameters in the SDSS $r$-band for the model indicated by FLAG$\_$FIT. Since we require accurate photometry and kinematics, we remove from our sample galaxies with FLAG$\_$FIT=3 in MPP-VAC (i.e., no available photometric parameters), as well as galaxies with unreliable spectra due to contamination by neighbours (removed after visual inspection).
When  FLAG$\_$FIT = 0 -- i.e.,  no preference between S\'ersic or S\'ersic + Exponential fits -- we use the values returned by the latter. (As discussed in Section~\ref{sec:FP*} we use S\'ersic rather than S\'ersic + Exponential parameters when computing the FP$_*$, since $k_n$ depends on the S\'ersic index, but our results do not significantly depend on the fit used.)

The MPP-VAC provides two estimates of the total magnitudes and sizes:
We always use the `truncated' magnitudes and sizes.  We do not circularize the sizes using the square root of the ratio of the semi-minor to semi-major axis, $\sqrt{b/a}$, as this would not be quite correct for S0s.  Instead, we treat $b/a$ as an additional parameter when fitting the FP.  Finally, we add $1.3z$ to the measured magnitudes to account for passive luminosity evolution. 

The MDLM-VAC provides morphological properties derived from supervised Deep Learning algorithms based on Convolutional Neural Networks following \citet{DS2018}.  To select early-types (as opposed to late-types) we require 
$$
 {\rm TType}\le 0,
$$
and to select S0s from these, we require
$$
P_{\rm S0} >  0.5
$$
as well.  
We subdivide the ETGs which are not S0s into slow and fast rotators (E-SR and E-FR) on the basis of the $\lambda_e-\epsilon$ diagnostic diagram, i.e. using the angular momentum $\lambda_e$ defined in \citet{Emsellem2007} (corrected for seeing following \citealt{Graham2018}) and the ellipticity $\epsilon$  (see Fig. 28--30 in \citealt[]{Fischer2019}).  Unless we specify otherwise, when we show results for E-FRs alone, then we require them to have $\lambda_e>0.2$.  However, when we combine E-SRs and E-FRs to make a sample of Es, then we include all values of $\lambda_e$.  For the discussion which follows, we will use E+S0 to refer to the full ETG sample (the E in E+S0 stands for both E-SRs and E-FRs).

\begin{figure*}
  \centering
  \includegraphics[width=0.8\linewidth]{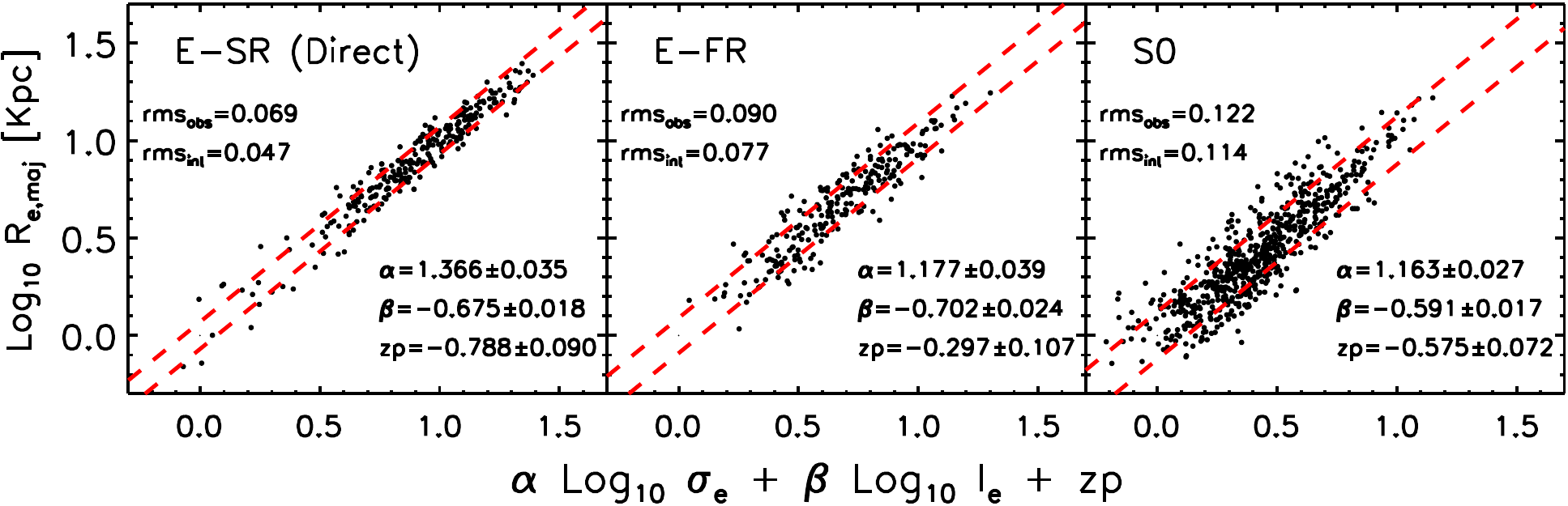}
  \includegraphics[width=0.55\linewidth]{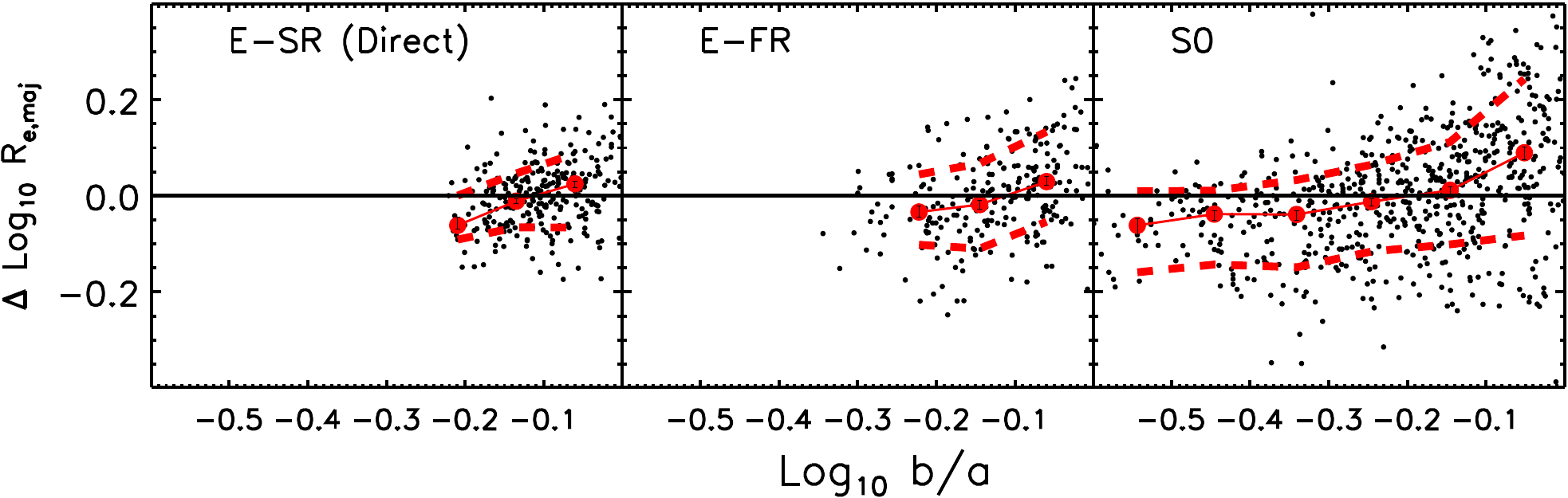}
  \includegraphics[width=0.8\linewidth]{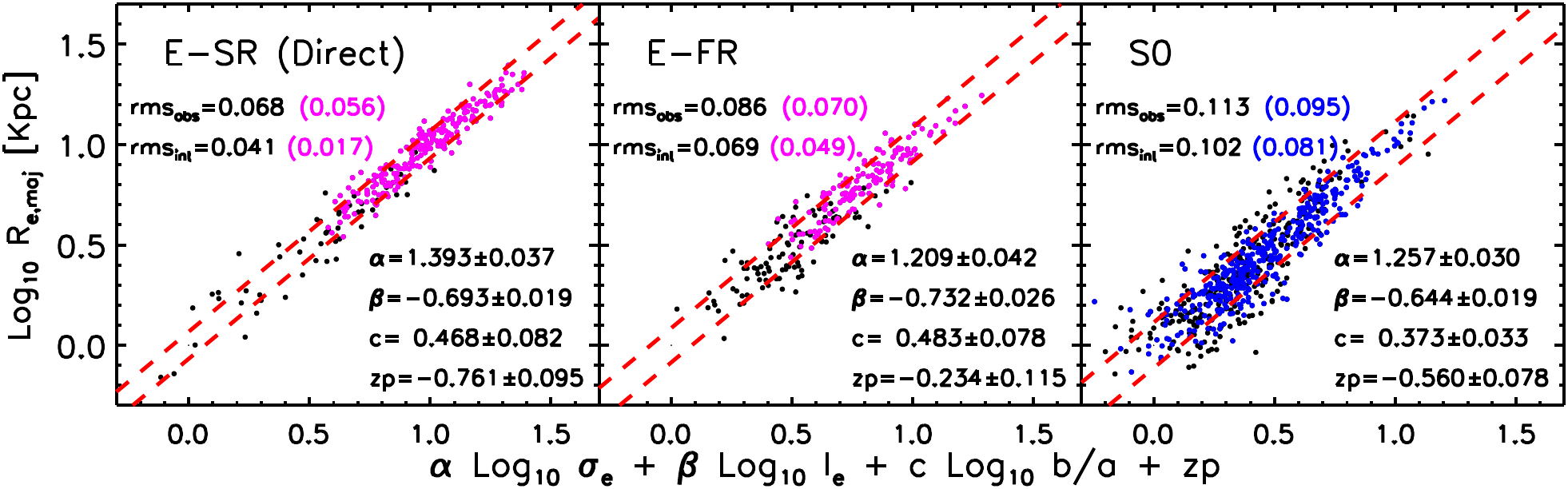}
  \caption{Dependence of FP$_{\rm L}$ on morphological type:  E-SRs (left), E-FRs (middle) and S0s (right).  Dashed lines show the rms scatter, rms$_{\rm obs}$.  Top: FP$_{\rm L}$ when $c=0$.  Middle:  Residuals from this FP$_{\rm L}$ correlate with image axis ratios $b/a$. Bottom:  FP$_{\rm L}$ when $c$ is determined by the fitting.  Top and bottom panels give the coefficients of the best-fitting FP$_{\rm L}$ for the sample, as well as the measured and intrinsic thickness. Magenta dots and values (in brackets) report the rms scatter when objects with $M_*< 10^{11}M_\odot$ are removed. Blue dots and values (in brackets) show the rms scatter for S0s with $b/a < 0.7$. }\label{fig:FP}
\end{figure*}

\subsection{Velocity dispersion estimates}
The Fundamental Plane requires an estimate of the velocity dispersion for each galaxy averaged within the projected half light radius.  We do this by constructing a composite spectrum by coadding all of a galaxy's spaxels that lie within the ellipse of $R_{e,maj}$ (given the axis ratio $b/a$ of the galaxy), and then estimate $\sigma_e$ from this coadded spectrum.  
The resulting estimate of $\sigma_e$ should be close to traditional estimates in the literature.  Note, however, that the absorption lines in this coadded spectrum are slightly broadened by rotation, which is different for each spaxel.  As a result, the dispersion estimated from this spectrum includes a nontrivial contribution from rotation; this contamination will matter more for S0s than for E-SRs, of course.  We discuss how we attempted to understand this contamination in Appendix~\ref{sec:sigVrot}, as well as the alternative procedure of shifting all spectra to restframe (thus removing rotation before stacking).  

\section{Results}\label{sec:results}
We describe the FP in two ways, which are known as the Direct and Orthogonal fits.

\subsection{The direct fit to the FP$_{\rm L}$}\label{sec:directFP}
Conceptually, the Direct fit is obtained by finding that set of $\alpha,\beta,c$ and $zp$ which minimize 
\begin{equation}
  \chi^2\equiv \sum_{i=1}^N \frac{(\log R_{e,maj} - \alpha \log \sigma_e - \beta \log I_e - c\,\log_{10}(b/a) - zp)^2}{N},
  \label{eq:FPe}
\end{equation}
where the sum is over all the $N$ objects in the sample.
As this returns the linear combination of $\log \sigma_e$, $\log I_e$ and $\log_{10}(b/a)$ which best predicts $\log R_{e,maj}$, it is the quantity of most interest in cosmological studies.
The thickness of the Plane is $\sqrt{\chi_{\rm min}^2}$; it quantifies the precision with which $\log R_{e,maj}$ is predicted.  

In practice, because of measurement errors, things are more complicated.  In the analysis below we account for measurement errors exactly as described in \cite{fp2012}.  This means that the $(\alpha,\beta,c)$ which define the FP are nearly but not exactly given by minimizing equation~(\ref{eq:FPe}).  Measurement errors also mean that it is important to distinguish between the intrinsic scatter around the FP, rms$_{\rm int}$, and the observed scatter, rms$_{\rm obs}$, which is larger because of measurement errors.  
Whereas rms$_{\rm obs}$ is the value of $\chi^2$ when the values $(\alpha,\beta,c)$ which define the FP are inserted in equation~(\ref{eq:FP}), the quantity rms$_{\rm int}$ is given by equation~(24) of \cite{fp2012}.

The top panels in Figure~\ref{fig:FP} show the FP$_{\rm L}$ defined by E-SRs, E-FRs and S0s (left to right), when $\sigma_e$ is close to the traditional one (it is estimated from a spectrum which is slightly broadened by rotation) and we force $c=0$ (i.e. we do not account for image ellipticity).  Each panel also provides the parameters $\alpha$ and $\beta$ of the FP$_{\rm L}$ in it, as well as rms$_{\rm obs}$ and rms$_{\rm int}$ which quantifies the tightness of the FP$_{\rm L}$.

There is an obvious dependence on morphology:  E-SRs define a steeper and significantly tighter FP$_{\rm L}$ than S0s, although some of these trends with morphology are simply due to the fact that different morphologies span different ranges of $L$ and $\sigma$, and the coefficients and scatter of the FP change if some $L$ or $\sigma$ are excluded \citep{HB2009}.  

The middle panels show that residuals from these FP$_{\rm L}$s correlate slightly with axis ratio $b/a$, indicating that we would do better if we allow $c$ to be determined by the fitting.  The bottom panels in Figure~\ref{fig:FP} show the results.  Comparison of the top and bottom panels shows that accounting for $b/a$ increases $\alpha$ and $\beta$ and reduces the scatter slightly.  Notice that $c\approx 0.45$.  Some authors work with `circularized' sizes, i.e. $R_{\rm e,maj}\sqrt{b/a}$ \cite[e.g.][]{comaFP,sdssFP,HB2009,Cappellari2013a}, which we noted in the Introduction would be inappropriate for S0s.  If this were the only dependence on $b/a$ then our analysis should have returned $c=-0.5-\beta$; this is about 0.25 smaller than the values we find. This is qualitatively consistent with results in \cite{comaFP}; the top right panel of their Figure~4 shows that even after circularizing $R_e$, residuals from the FP defined by 61 Coma cluster galaxies correlates with $b/a$. 

The middle panels of Figure~\ref{fig:FP} (especially the S0 panel) show that the scatter is larger for objects with $b/a > 0.7$ ($\log_{10}b/a > -0.15$).  There is a plausible physical explanation:  these are nearly face-on so, at least for S0s, the rotational contribution to the total energy is systematically underestimated \cite[see, e.g.][for more extensive discussion of inclination effects]{comaFP}.  While allowing $c\ne 0$ removes the mean trends shown in these middle panels, the increased scatter for face-on objects remains.  The numbers in blue brackets in the bottom right panel of Figure~\ref{fig:FP} show the scatter around FP$_{\rm L}$ if objects with $b/a > 0.7$ are removed from the sample (after fitting):  this reduces the rms scatter by 0.02~dex (approximately 5\%).  

\begin{figure*}
  \centering
  \includegraphics[width=0.85\linewidth]{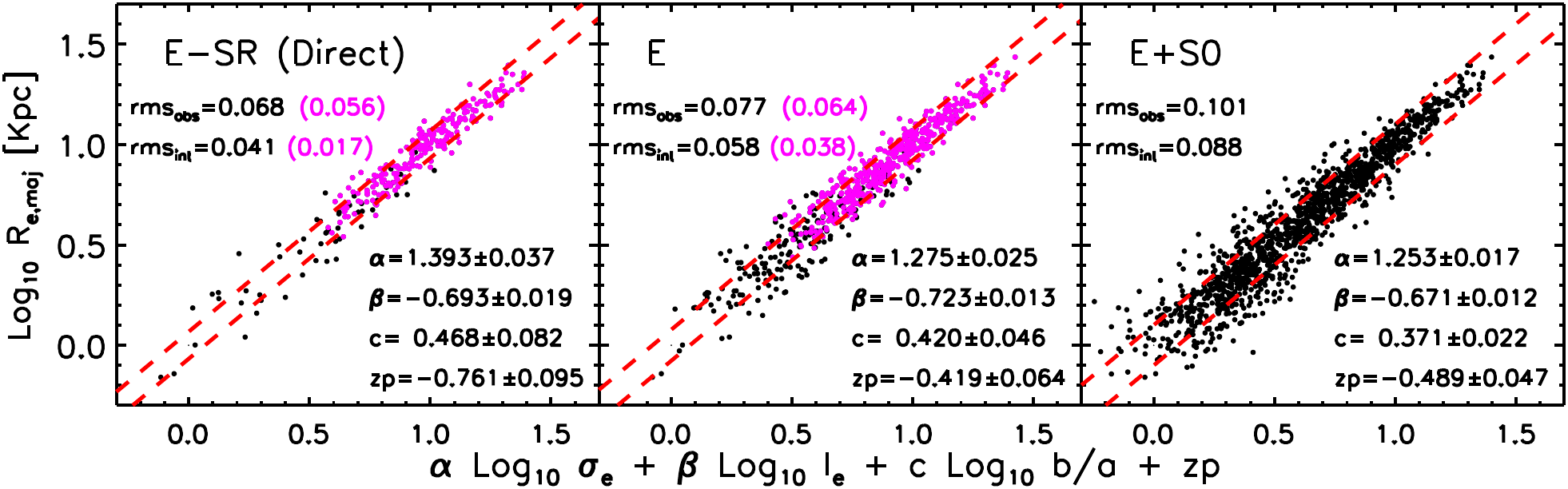}
  \caption{`Cumulative' version of FP$_{\rm L}$ shown in bottom panels of Figure~\ref{fig:FP}:  middle panels show Es = E-SR + E-FRs, and right hand panels show E+S0s. }
  \label{fig:FPeall}
\end{figure*}

\begin{figure*}
  \centering
  \includegraphics[width=0.85\linewidth]{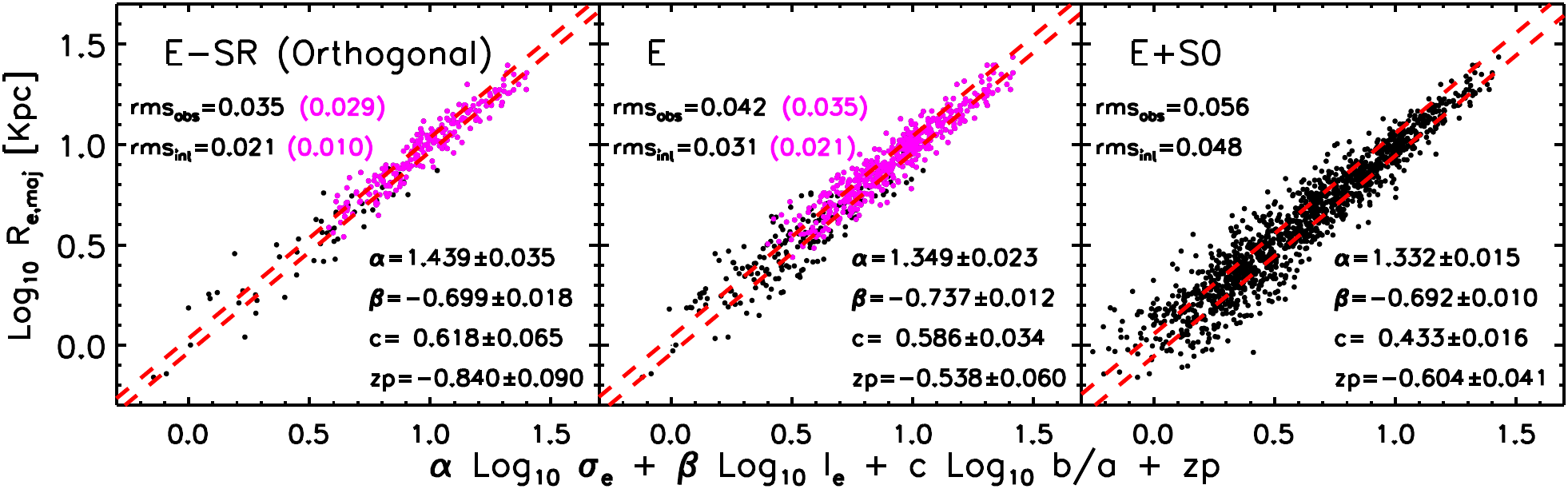}
  \caption{Same as Figure~\ref{fig:FPeall}, but now showing the orthogonal fits.  }
  \label{fig:orthFPe}
\end{figure*}

Figure~\ref{fig:FPeall} shows cumulative versions of the FP$_{\rm L}$s in the bottom panel of Figure~\ref{fig:FP}.  Left to right panels show E-SRs (same as bottom left panel of Figure~\ref{fig:FP}), Es = E-SRs + E-FRs (here we also include E-FRs with $\lambda_e<0.2$) and Es+S0s.
The right panel is closest to most FP determinations in the literature, and the observed rms is similar to much previous work \citep{sdssFP,HB2009,Cappellari2013a}.  The scatter is reduced substantially if S0s are excluded (from $\sim 23\%$ to $\sim 18\%$); this can be done fairly easily even without IFU data (the MDLM-VAC morphological classifications only require photometry).  Objects with $M_*< 10^{11}M_\odot$ contribute significantly to the scatter, presumably because these are the objects for which inclination and rotation matter more (see Appendix~\ref{sec:sigVrot} for discussion of the uncertainties due to these effects).  The magenta numbers show that when they are removed, then the observed scatter is even further reduced: A sample of massive Es (which can be selected without IFU data) can provide distance determinations with rms errors of just $15\%$ (potentially $9\%$).

If E-FRs are excluded from the full sample of Es, then the observed scatter of 0.068~dex means that E-SRs can provide a 16\% distance determination; the intrinsic scatter of just 0.041~dex suggests this can be reduced to 9\%.  Of course, this requires IFU data or measurements using a slit. Massive E-SRs, with observed scatter of $\sim 0.056$~dex (intrinsic $0.017$~dex), can provide 13\% (potentially $\sim 4\%$) distance determinations. 

\begin{figure}
  \centering
  \includegraphics[width=0.8\linewidth]{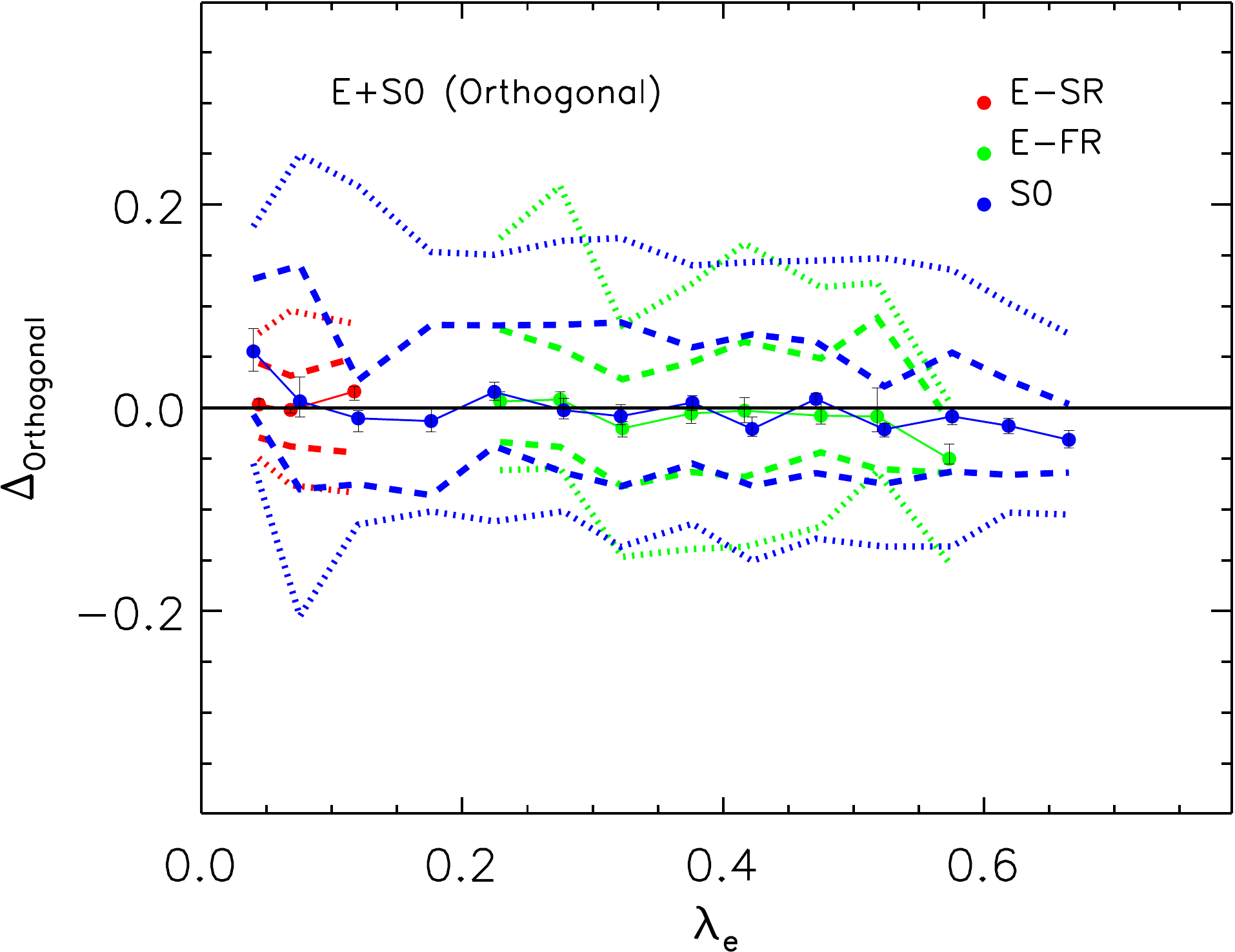}
  \caption{When subdivided by morphological type, residuals from the orthogonal fit to the FP$_{\rm L}$ of E+S0s do not correlate with $\lambda_e$, even though the FP$_{\rm L}$s for each morphological type are different (e.g. Figure~\ref{fig:FP}).  Dashed and dotted lines show the regions which enclose 68\% and 95\% of the objects in each $\lambda_e$-bin. }
  \label{fig:residLMs}
\end{figure}

\subsection{The orthogonal fit to the FP$_{\rm L}$}\label{sec:orthFP}
The orthogonal fit coefficients $\alpha$, $\beta$ and $c$ describe the eigenvector associated with the smallest eigenvalue of the $4\times 4$ symmetric matrix whose elements are the error-corrected covariances between the four observables $\log R_{e,maj}$, $\log \sigma_e$, $\log I_e$ and $\log_{10}(b/a)$.  The intrinsic scatter is the square-root of this eigenvalue, whereas the `observed' scatter is given by evaluating $\chi^2$ of equation~(\ref{eq:FPe}) using the orthogonal fit coefficients, dividing by $(1 + \alpha^2 + \beta^2 + c^2)$, and taking the square-root.  These are the coefficients which are more closely related to the `tilt' of the FP.  

\begin{figure*}
  \centering
  \includegraphics[width=0.9\linewidth]{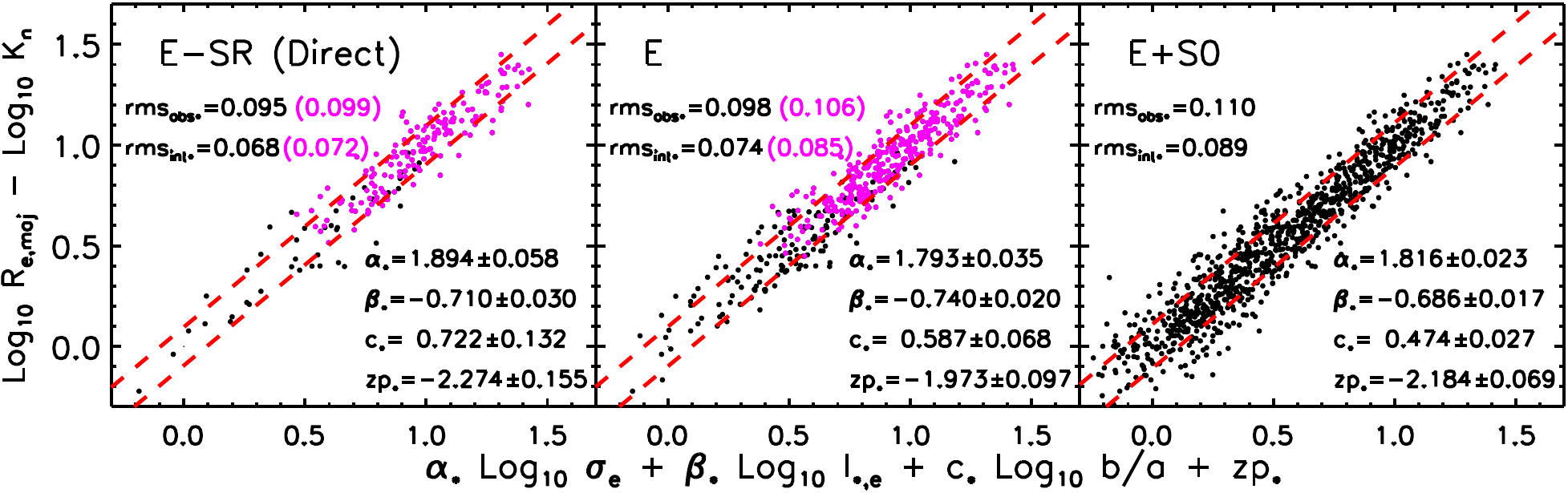}
  \includegraphics[width=0.9\linewidth]{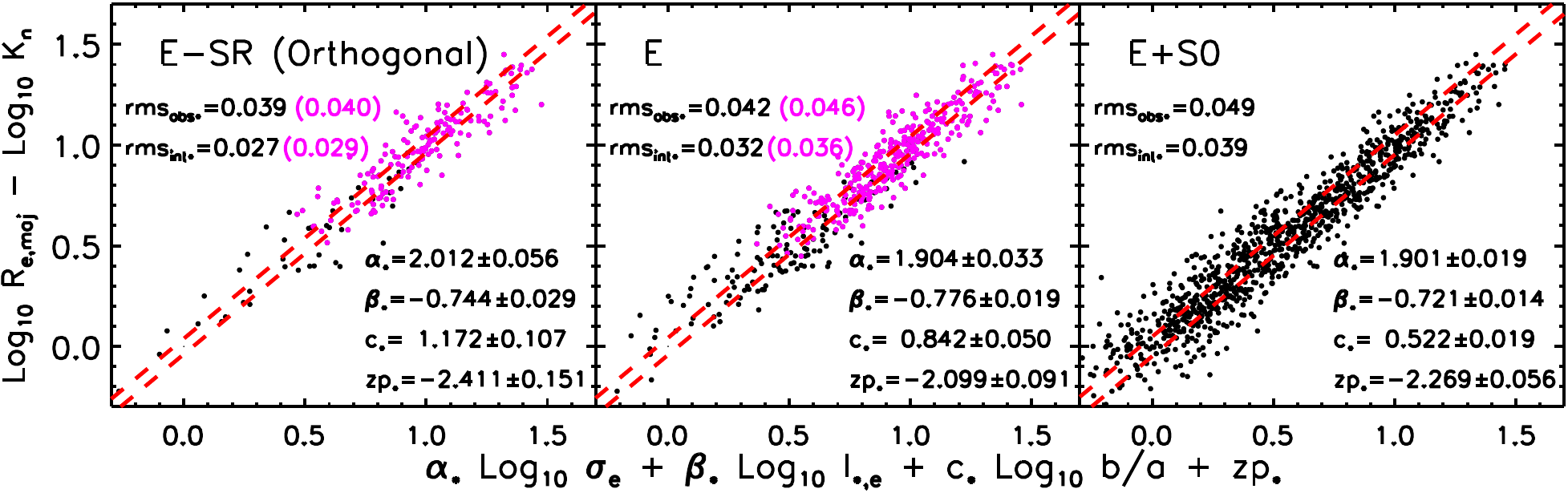}  \caption{Direct and orthogonal fits to FP$_{*}$, the stellar mass Fundamental Plane. The measured values in both the x- and y-axis were shifted by 0.7 to maintain the same range as in the previous figures.  Magenta dots and values (in brackets) report the rms scatter when objects with $M_*< 10^{11}M_\odot$ are removed.}
  \label{fig:FP*}
\end{figure*}

Figure~\ref{fig:orthFPe} is similar to Figure~\ref{fig:FPeall}, but now showing the orthogonal fit.  While the qualitative trends with morphology are similar to those for the direct fit, the coefficient $\alpha$ is larger and the scatter smaller than for the direct fit, for reasons discussed in \cite{fp2012}.  Note in particular that the intrinsic scatter around the orthogonal FP is just 5\% for E-SRs, and half this value if we only select objects with $M_*>10^{11}M_\odot$.  This is what has motivated the title of this paper.

Figure~\ref{fig:residLMs} makes the point that E-SRs define a thin plane in a slightly different way.  It shows residuals from the orthogonal fit shown in the right-hand panel of Figure~\ref{fig:orthFPe} -- i.e. the FP$_{\rm L}$ defined by the full sample of E+S0s -- as a function of $\lambda_e$ subdivided by morphological type.  While there is no correlation for any of the three subsamples, it is clear that E-SRs show smaller scatter.

We have also looked for correlations between residuals from these FP$_{\rm L}$s and stellar mass and found none.  However, for Es, the scatter increases at low masses ($M_*\le 10^{11}M_\odot$).  Removing low mass objects reduces the thickness of FP$_{\rm L}$ substantially (these are the numbers reported in magenta in the previous figures).

\subsection{ The FP based on stellar mass: FP$_*$}\label{sec:FP*}
The orthogonal fit coefficients in Figure~\ref{fig:orthFPe} are far from the virial values.  Equation~(\ref{eq:FP}) suggests that this might be due to variations in $k_n$ and $M_*/L$ across the populations.  

To account for variations in the stellar mass to light ratio,  we need $M_*/L$ for each object.  We take these from \cite{Mendel2014}.  In practice, they provide $M_*$ values which were obtained by multiplying their reliable $M_*/L$ estimates by estimates of $L$ known as SDSS {\tt model} magnitudes.  Therefore, we take their published $M_*$ values and divide them by the SDSS {\tt model} $L$ values they used.  By only using their $M_*/L$ values we are ensuring that the photometric parameters which enter our FP$_*$ determinations remain those of {\tt PyMorph}, which are more reliable \citep{F2017,Bernardi2017a}.  We use these to define $I_{*e}\equiv (M_*/L)\,I_e$. 

\begin{figure*}
  \centering
  \includegraphics[width=0.8\linewidth]{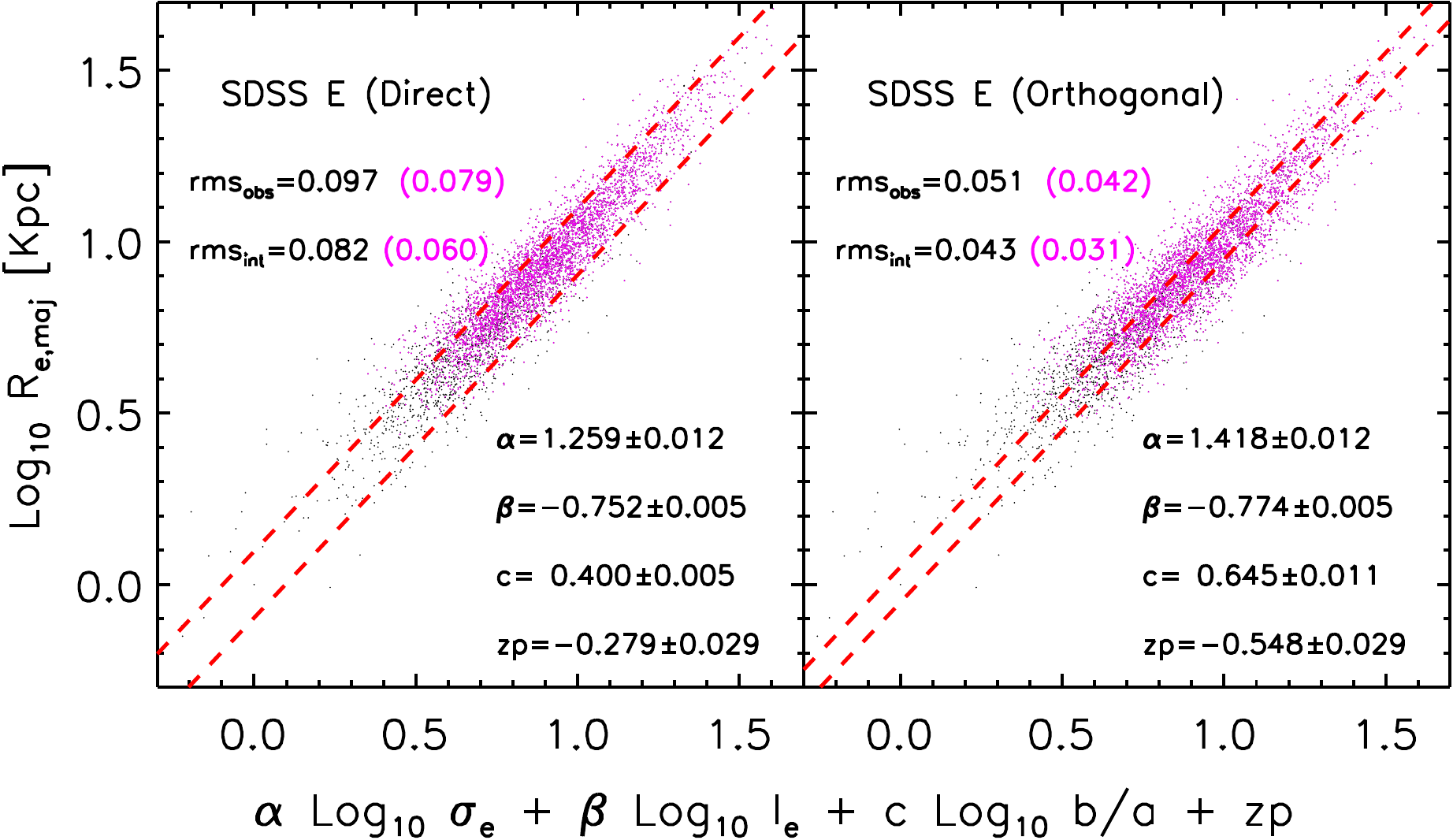}
  \includegraphics[width=0.8\linewidth]{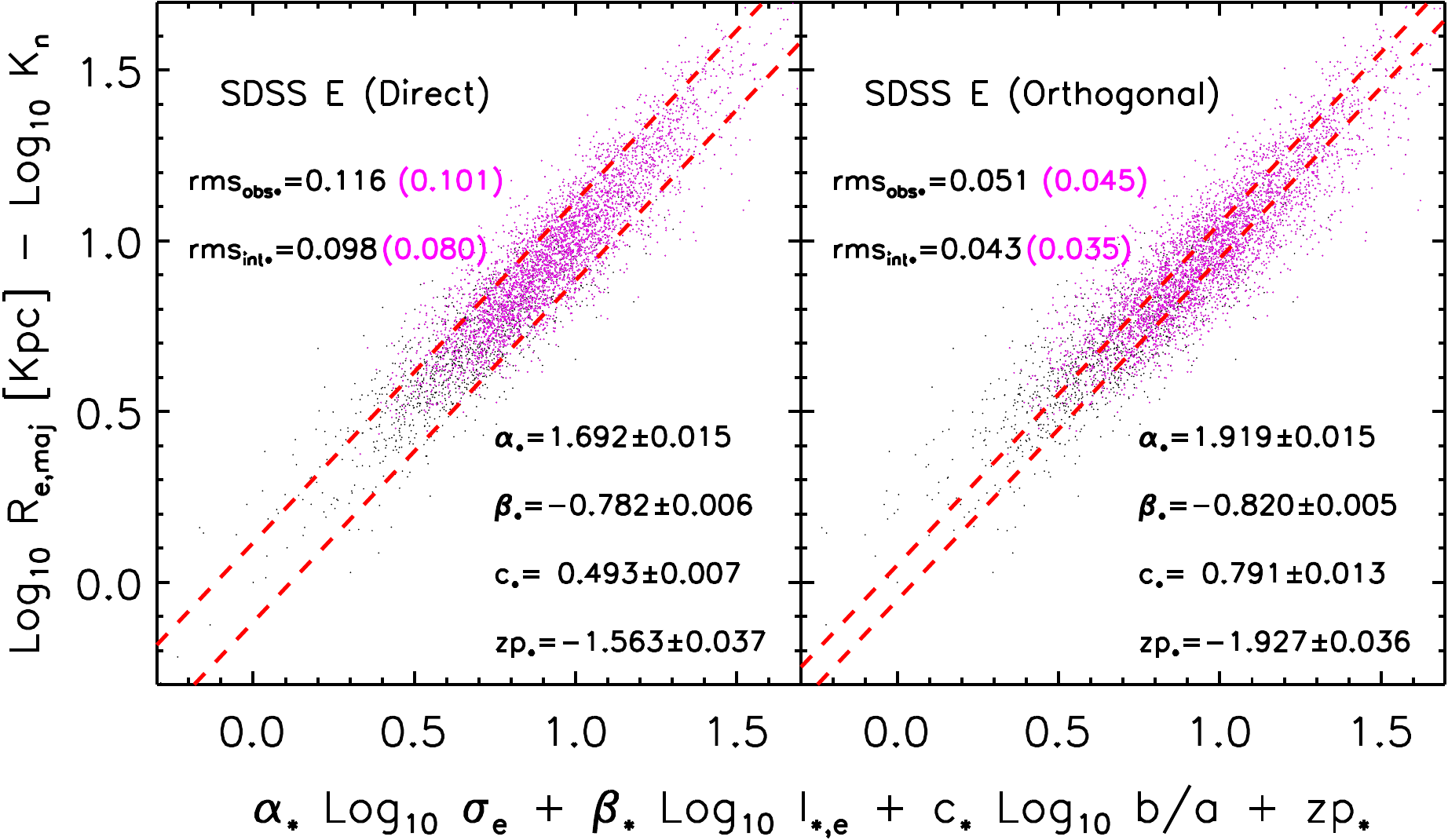}
  \caption{Top:  Direct (left) and orthogonal (right) fits to FP$_{\rm L}$ in the SDSS.  Black dots show 5000 of the SDSS Es (selected randomly); magenta dots show the subset of these that have $M_*\ge 10^{11}M_\odot$; excluding lower mass objects reduces the scatter.  Bottom:  Same as top, but for FP$_*$; the measured values in both the x- and y-axis were shifted by 0.7 to maintain the same range as in the top panel. }
  \label{fig:sdssFP}
\end{figure*}

The {\tt PyMorph} photometry is based on S\'ersic or S\'ersic+Exponential fits to the surface brightness.  If $M_*/L$ gradients can be ignored, then the shape of the light profile is simply related to that of $M_*$.  In addition, if the stellar mass dominates on the scales which determine $\sigma_e$, then, for the objects which are well fit by single S\'ersic profiles, we can take $k_n$ from Table~1 of \citet[][also see their Figure~1]{Bernardi2018a}, where $n$ is the S\'ersic index.  We use the values of $k_n$ that are appropriate for an aperture of size $R_e$ because that is the scale used for $\sigma_e$ in the previous section -- and is the scale usually used for FP analyses -- even though it is unlikely that the mass within $R_e$ is dominated by the stars \citep{Bernardi2018b}.

Since the $k_n$ values we would like to use are not appropriate for S\'ersic+Exponential profiles, the FP$_*$ results which follow are based on the subset of objects for which a single S\'ersic profile is a good fit.  We have checked that this subset defines the same FPs as shown previously for the full sample -- i.e., the additional requirement on profile shape does not change any of our conclusions about how FP$_{\rm L}$ depends on morphological type.

Figure~\ref{fig:FP*} shows direct and orthogonal fits to FP$_{*}$ which result from replacing $I_e\to I_{*e}$ and $R_{e,maj}\to R_{e,maj}/k_n$ and then repeating the analysis which lead to the FPs of the previous Section.  The observed scatter around FP$_*$ is somewhat larger than for FP$_{\rm L}$; presumably this is because the stellar mass estimates are noisier than the luminosities themselves.  Notice that now $\alpha_* \approx 2$ for all the orthogonal fits.
However, the $\beta_*\approx 0.75$ values we find differ from the $-1$ expected from the virial theorem.  To be consistent with equation~(\ref{eq:FP}), our FP$_*$ coefficients suggest that $M_*/M\propto I_{*,e}^{1+\beta_*}\propto I_{*,e}^{0.25}$.  Equation~(\ref{eq:M*Mtot}) shows that this ratio depends on the product of $M_*/M_{\rm bar}$ with $(1-f_{\rm DM})$.  Therefore, if $f_{\rm DM}$ is constant across the population, then $M_*/M_{\rm bar}$ increases with $I_{*e}$.  This could result from the correlation between star formation rate and gas surface density in their disk dominated progenitors.  Alternatively, if $M_*/M_{\rm bar}$ is constant across the population, then the dark matter fraction in Es decreases as $I_{*e}$ increases; this is also physically reasonable. 

With these interesting implications for $M_*/M_{\rm bar}$ and $f_{\rm DM}$ in mind, it is worth recalling that $k_n$ assumes that $M_*/L$ gradients can be ignored.  Figure~16 of \cite{BDS2019} suggests that gradients matter in this context, so they may modify the values of $\alpha_*$ and $\beta_*$.  Unfortunately, as they driven by IMF gradients, they are not available on an object-by-object basis, so we cannot self-consistently account for their effects here.

%

In a recent analysis of MaNGA early-type galaxies, \cite{Li2018} reported good agreement between a quantity they called $M_{1/2}$, which is the enclosed mass (stars + gas + dark matter) within a spherical radius $r_{1/2}$ determined from their best-fitting Jeans Anisotropic Models, and $R_{e,maj}^{0.96}\sigma_e^{1.96}$.  (Their analysis ignores $M_*/L$ gradients.)  
Since their $M_{1/2}$ explicitly involves the Jeans equation, this agreement with the virial scaling (compare our equation~\ref{eq:Mdyn}) is not so surprising.  In fact, if one sets measurement errors to zero, then the orthogonal fit to the plane defined by the $M_{1/2}$, $R_{e,maj}$ and $\sigma_e$ values provided in their Table~A1 for Es has the virial coefficients with scatter that is smaller than 0.02~dex.  As this is smaller than the measurement errors they quote, this strongly suggests they are essentially comparing one quantity with itself.  In contrast, our FP$_*$ analysis is less directly tied to the virial theorem: whereas $k_n$ comes from the shape of the light profile (and one may view $k_n$ as coming from the Jeans equation) our $M_*/L$ comes from analysis of the stellar populations -- not the dynamics.

\section{The SDSS FP and comparison with MaNGA}\label{sec:sdss}
The MaNGA selection function is complex \cite[see][]{Wake2017}.
Since cuts in $L$ and $\sigma$ can bias the coefficients of the FP \citep{sdssFP,HB2009}, it is useful to see how the MaNGA FP compares with that in a sample which is selected differently.  For this, we have chosen the SDSS, which is magnitude-limited, so selection effects can be removed \cite[e.g.][]{fp2012}.  To simplify the comparison, we use photometric parameters which are based on the same {\tt PyMorph} algorithm that was used to produce MPP-VAC, and morphological information from the same DL which provided the MDLM-VAC classifications.  These are available from \cite{Meert2015} and \cite{DS2018}, respectively, for a sample of $\sim 670,000$ SDSS galaxies.  From these, we select Es by requiring ${\rm TType}\le 0 $ and $P_{\rm S0} <  0.5$, and that the S\'ersic+Exponential fit has {\tt finalflag = 2} (this indicates that it is single-component bulge).  This leaves us with a sample of about 10\% of the original sample.  (Note that we cannot distinguish between E-SR and E-FRs because we do not have IFU data for these objects.) When we study FP$_*$, we also require $M_*/L$ estimates from \cite{Mendel2014}.

The top panel of Figure~\ref{fig:sdssFP} shows direct and orthogonal fits to FP$_{\rm L}$ in the SDSS \cite[we account for selection effects and measurement errors following][]{fp2012}.  The coefficients $\alpha$ are similar to those in the middle panels of Figures~\ref{fig:FPeall} and~\ref{fig:orthFPe} (i.e. for E-SR + E-FRs), whereas the coefficients $\beta$ are slightly (0.05~dex) larger (i.e., more negative).  The zero-points are quite different, but it is well known that these are particularly sensitive to selection effects.  The SDSS scatter is slightly larger, but if we restrict to masses greater than $M_* > 10^{11}M_\odot$, then the scatter is reduced and the agreement with MaNGA improves.  This suggests that the complex MaNGA selection does not severely affect our conclusions about the thickness of FP$_{\rm L}$.  

In addition to the agreement of the FP$_{\rm L}$ coefficients which describe the direction of the vector orthogonal to the plane, the MaNGA and SDSS samples are also in reasonable agreement about the eigenvectors within the plane.  \cite{efarVI} noted that, in their sample, one of the vectors in the plane has essentially no component in the $\log\sigma$ direction.  This is also true of the orthogonal FP in the SDSS reported by \cite{HB2009}, and remains true with the newer {\tt PyMorph} photometry, for which this vector has components $(1,-0.9,0.2)$ in the $(\log_{10}R_e,\log_{10}I_i,0.2\log_{10}\sigma_e)$ directions.  However, for MaNGA Es (i.e. E-SR + E-FRs) the coefficient in the $I_e$ direction is the same, and that in the $\sigma_e$ direction is 0.25.  Following \cite{fp2012}, we argue that this is primarily a consequence of the fact that the distribution of $I$ is much wider than that of $\sigma$, and $I$ and $\sigma$ are almost uncorrelated. 

Finally, the bottom panel in Figure~\ref{fig:sdssFP} shows the corresponding stellar mass FP$_*$s.  These have larger $\alpha_*$, more negative $\beta_*$ and similar (or slightly larger) scatter, just as in MaNGA.  For the SDSS, we have explored using the measured velocity dispersion $\sigma_{\rm fiber}$, rather than extrapolating the measured value to $\sigma_e$.  In this case, provided we use the appropriate $k_n$ term from \cite{Bernardi2018a}, this change makes little difference to our results.  

\begin{table*}
\centering
MaNGA-FP$_{\rm L}$: MaNGA LUMINOSITY FUNDAMENTAL PLANE\\ 
 \begin{tabular}{l c c c c c c c c c c}
 \hline  
 Direct & $\alpha$ & $\beta$ & $c$ & zp & rms$_{\rm obs}$ & rms$_{\rm int}$ & (rms$_{\rm obs}$) & (rms$_{\rm int}$) \\ 
 \hline
                      E-SR & $ 1.393 \pm  0.037$ & $-0.693 \pm  0.019$ & $ 0.468 \pm  0.082$ & $-0.761 \pm  0.095$ & $0.068$ & $0.041 $ & ($0.056$) & ($0.017$) \\
E-FR ($\lambda_e < 0.2$) & $ 1.229 \pm  0.057$ & $-0.749 \pm  0.027$ & $ 0.374 \pm  0.089$ & $-0.255 \pm  0.146$ & $0.075$ & $0.055 $ & ($0.067$) & ($0.044$) \\
E-FR ($\lambda_e > 0.2$) & $ 1.209 \pm  0.042$ & $-0.732 \pm  0.026$ & $ 0.483 \pm  0.078$ & $-0.234 \pm  0.115$ & $0.086$ & $0.069 $  & ($0.070$) & ($0.049$) \\
                      S0 & $ 1.257 \pm  0.030$ & $-0.644 \pm  0.019$ & $ 0.373 \pm  0.033$ & $-0.560 \pm  0.078$ & $0.113$ & $0.102 $  & & \\

 \hline
                       E & $ 1.275 \pm  0.025$ & $-0.723 \pm  0.013$ & $ 0.420 \pm  0.046$ & $-0.419 \pm  0.064$ & $0.077$ & $0.058 $ & ($0.064$) & ($0.038$) \\
                    E+S0 & $ 1.253 \pm  0.017$ & $-0.671 \pm  0.012$ & $ 0.371 \pm  0.022$ & $-0.489 \pm  0.047$ & $0.101$ & $0.088 $ & & \\
 
 \hline
 \hline
 Orthogonal & $\alpha$ & $\beta$ & $c$ & zp & rms$_{\rm obs}$ & rms$_{\rm int}$ \\ 
 \hline
                    E-SR & $ 1.439 \pm  0.035$ & $-0.699 \pm  0.018$ & $ 0.618 \pm  0.065$ & $-0.840 \pm  0.090$ & $0.035$ & $0.021 $ & ($0.029$) & ($0.010$) \\
E-FR ($\lambda_e < 0.2$) & $ 1.313 \pm  0.054$ & $-0.771 \pm  0.025$ & $ 0.511 \pm  0.059$ & $-0.382 \pm  0.137$ & $0.041$ & $0.030 $ & ($0.038$) & ($0.025$) \\
E-FR ($\lambda_e > 0.2$) & $ 1.308 \pm  0.039$ & $-0.773 \pm  0.024$ & $ 0.739 \pm  0.058$ & $-0.327 \pm  0.108$ & $0.046$ & $0.037 $ & ($0.039$) & ($0.028$) \\
                      S0 & $ 1.450 \pm  0.027$ & $-0.721 \pm  0.017$ & $ 0.541 \pm  0.023$ & $-0.740 \pm  0.072$ & $0.061$ & $0.055 $ & & \\

 \hline
                       E & $ 1.349 \pm  0.023$ & $-0.737 \pm  0.012$ & $ 0.586 \pm  0.034$ & $-0.538 \pm  0.060$ & $0.042$ & $0.031 $ & ($0.035$) & ($0.021$) \\
                    E+S0 & $ 1.332 \pm  0.015$ & $-0.692 \pm  0.010$ & $ 0.433 \pm  0.016$ & $-0.604 \pm  0.041$ & $0.056$ & $0.048 $ & & \\
 \hline
 \hline
 \end{tabular}
 \caption{Direct and Orthogonal fit coefficients of the FP$_{\rm L}$ for various morphological samples, and associated measured and intrinsic scatter, and for the subset having $M_*\ge 10^{11}M_\odot$ (in brackets).}
 \label{tab:FP}
\end{table*}

\begin{table*}
\centering
MaNGA-FP$_{*}$: MaNGA STELLAR MASS FUNDAMENTAL PLANE\\
\begin{tabular}{l c c c c c c c c}
 \hline  
 Direct & $\alpha_*$ & $\beta_*$ & $c_*$ & zp$_*$ & rms$_{\rm obs*}$ & rms$_{\rm int*}$ & (rms$_{\rm obs*}$) & (rms$_{\rm int*}$) \\ 
 \hline
                     E-SR & $ 1.894 \pm  0.058$ & $-0.710 \pm  0.030$ & $ 0.722 \pm  0.132$ & $-2.274 \pm  0.155$ & $0.095$ & $0.068 $ & ($0.099$) & ($0.072$) \\
E-FR ($\lambda_e < 0.2$) & $ 1.851 \pm  0.088$ & $-0.820 \pm  0.047$ & $ 0.693 \pm  0.153$ & $-1.877 \pm  0.239$ & $0.105$ & $0.080 $ & ($0.105$) & ($0.079$) \\
E-FR ($\lambda_e > 0.2$) & $ 1.748 \pm  0.062$ & $-0.759 \pm  0.043$ & $ 0.568 \pm  0.104$ & $-1.815 \pm  0.184$ & $0.095$ & $0.070 $ & ($0.093$) & ($0.068$) \\
                      S0 & $ 1.800 \pm  0.040$ & $-0.644 \pm  0.026$ & $ 0.430 \pm  0.035$ & $-2.278 \pm  0.109$ & $0.112$ & $0.093 $ & &\\

 \hline

                        E & $ 1.793 \pm  0.035$ & $-0.740 \pm  0.020$ & $ 0.587 \pm  0.068$ & $-1.973 \pm  0.097$ & $0.098$ & $0.074 $ & ($0.106$) & ($0.085$) \\
                    E+S0 & $ 1.816 \pm  0.023$ & $-0.686 \pm  0.017$ & $ 0.474 \pm  0.027$ & $-2.184 \pm  0.069$ & $0.110$ & $0.089 $ & & \\
 
 \hline
 \hline
 
 Orthogonal & $\alpha_*$ & $\beta_*$ & $c_*$ & zp$_*$ & rms$_{\rm obs*}$ & rms$_{\rm int*}$ & (rms$_{\rm obs*}$) & (rms$_{\rm int*}$) \\ 
 \hline
                    E-SR & $ 2.012 \pm  0.056$ & $-0.744 \pm  0.029$ & $ 1.172 \pm  0.107$ & $-2.411 \pm  0.151$ & $0.039$ & $0.027 $ & ($0.040$) & ($0.029$) \\
E-FR ($\lambda_e < 0.2$) & $ 2.024 \pm  0.082$ & $-0.901 \pm  0.044$ & $ 1.056 \pm  0.096$ & $-2.016 \pm  0.224$ & $0.043$ & $0.032 $ & ($0.044$) & ($0.034$) \\
E-FR ($\lambda_e > 0.2$) & $ 1.865 \pm  0.057$ & $-0.820 \pm  0.040$ & $ 0.760 \pm  0.077$ & $-1.886 \pm  0.169$ & $0.042$ & $0.031 $ & ($0.040$) & ($0.029$) \\
                      S0 & $ 1.993 \pm  0.035$ & $-0.728 \pm  0.022$ & $ 0.539 \pm  0.025$ & $-2.427 \pm  0.095$ & $0.050$ & $0.040 $ & &\\

 \hline
                    E & $ 1.904 \pm  0.033$ & $-0.776 \pm  0.019$ & $ 0.842 \pm  0.050$ & $-2.099 \pm  0.091$ & $0.042$ & $0.032 $ & ($0.046$) & ($0.036$) \\
                    E+S0 & $ 1.901 \pm  0.019$ & $-0.721 \pm  0.014$ & $ 0.522 \pm  0.019$ & $-2.269 \pm  0.056$ & $0.049$ & $0.039 $ & &\\

 \hline
 \end{tabular}
 \caption{Direct and Orthogonal fit coefficients of the stellar mass Plane FP$_{*}$ for various morphological samples, and associated measured and intrinsic scatter, and for the subset having $M_*\ge 10^{11}M_\odot$ (in brackets).}
 \label{tab:FP*}
\end{table*}

\begin{table*}
\centering
SDSS FUNDAMENTAL PLANE\\ 
  \begin{tabular}{l c c c c c c c c}
 \hline  
 SDSS-FP$_{\rm L}$ & $\alpha$ & $\beta$ & $c$ & zp & rms$_{\rm obs}$ & rms$_{\rm int}$ & (rms$_{\rm obs}$) & (rms$_{\rm int}$) \\ 
 \hline
  Direct & $ 1.259 \pm  0.012$ & $-0.752 \pm  0.005$ & $ 0.400 \pm  0.005$ & $-0.279 \pm  0.029$ & $0.097  $ & $0.082 $ & ($0.079$) & ($0.060  $) \\
  Orthogonal & $ 1.418 \pm  0.012$ & $-0.774 \pm  0.005$ & $ 0.645 \pm  0.011$ & $-0.548 \pm  0.029$ & $0.051  $ & $0.043 $ & ($0.042$) & ($0.031  $) \\
 \hline
 \hline
 SDSS-FP$_{*}$ & $\alpha_*$ & $\beta_*$ & $c_*$ & zp$_*$ & rms$_{\rm obs*}$ & rms$_{\rm int*}$ & (rms$_{\rm obs*}$) & (rms$_{\rm int*}$)\\ 
 \hline
    Direct & $ 1.692 \pm  0.015$ & $-0.782 \pm  0.006$ & $ 0.493 \pm  0.007$ & $-1.563 \pm  0.037$ & $0.116  $ & $0.098 $ & ($0.101$) & ($0.080  $) \\
    Orthogonal & $ 1.919 \pm  0.015$ & $-0.820 \pm  0.005$ & $ 0.791 \pm  0.013$ & $-1.927 \pm  0.036$ & $0.051  $ & $0.043 $ & ($0.045$) & ($0.035  $) \\
 \hline
 \hline
 \end{tabular}
 \caption{Direct and Orthogonal fit coefficients of FP$_{\rm L}$ and FP$_*$ in the SDSS.  Observed and intrinsic scatter are given for the full sample, and for the subset having $M_*\ge 10^{11}M_\odot$ (in brackets).}
 \label{tab:sdssFP}
\end{table*}

\section{Conclusions}\label{sec:end}
We have measured the Fundamental Plane of MaNGA early-type galaxies.

The FP requires an estimate of the velocity dispersion within a standardized aperture $\sigma_e$ for each object.  As we discuss in Appendix~\ref{sec:sigVrot}, there is some freedom in how this can be done in IFU datasets:  e.g., one may or may not choose to shift spectra to restframe before stacking.  While our results are insensitive to this difference, tests in Appendix~\ref{sec:sigVrot} suggest that one should be cautious when using the estimated rotation and dispersion values in MaNGA spaxels (if one wants better than 10\% accuracy).

We have found that the MaNGA FP varies with morphological type (Figures~\ref{fig:FP}--\ref{fig:orthFPe}).  Our results are summarized in Table~\ref{tab:FP}.  They show that the FP of slow rotators is remarkably tight:  it potentially provides distances to $\sim 10\%$ accuracy, so should also prove useful in peculiar velocity and magnification bias studies.  Additionally, the fact that the scatter orthogonal to the E-SR plane is just $\sim 5\%$ provides a stringent constraint on models of how slow rotators formed.

When luminosity is replaced by stellar mass to make the stellar mass Plane FP$_*$ (Figure~\ref{fig:FP*}), then the coefficient of the velocity dispersion term becomes much closer to that expected from the virial theorem (Table~\ref{tab:FP*}). However, the coefficient of the stellar surface brightness term is different, suggesting that the ratio of dynamical to stellar mass varies with stellar surface brightness (equation~\ref{eq:FP}).  At face value, the variation suggests that the dark matter fraction is smaller and/or the ratio of stellar to baryonic mass is larger when $I_{*e}$ is larger (equation~\ref{eq:M*Mtot}).  However, we expect that self-consistently including $M_*/L$ gradients in $k_n$ and $I_{*e}$ of equation~(\ref{eq:FP*}) will modify the FP$_*$ coefficients, perhaps changing these conclusions.  

Despite the complex MaNGA selection rules, the intrinsic correlations in MaNGA are not very different from those in the full SDSS once the SDSS magnitude limit has been accounted for (Figure~\ref{fig:sdssFP}).  The SDSS FP and FP$_*$ parameters in Table~\ref{tab:sdssFP} supercede those of \cite{HB2009} as they are based on better $M_*/L$ ratios and better photometry (and, because the photometry is based on S\'ersic profiles, we now include $k_n$ in the virial scaling).  As for MaNGA, the coefficient of $\sigma$ in FP$_*$ of the SDSS is remarkably close to the virial scaling, whereas that for $I_{*e}$ is not.  Whether gradients will bring the FP$_*$ even closer to the virial theorem form remains to be seen.

Although discussion of the orthogonal fit to the FP usually focusses on the eigenvector and its length orthogonal to the Plane (for the orthogonal fit), Appendix~\ref{sec:FPnorm} discusses an additional constraint associated with the structure within the Plane.  Namely, when all variables have been normalized by their rms values, then the vector `across' the plane has almost no dependence on size and the coefficients of the normalized surface brightness and velocity dispersion axes are nearly equal.  This indicates that $I_e$ and $\sigma_e$ are almost independent, whereas the $R_e-I_e$ and $R_e-\sigma_e$ correlation coefficients are nearly equal and opposite \citep{fp2012}.  As this holds in both the SDSS and MaNGA, realistic models of early type galaxy formation should be able to reproduce this FP structure.  Indeed, the independence of $I_e$ and $\sigma_e$ provide the relevant background to recent work arguing that $I_e$ and $\sigma_e$ are fundamental for understanding the stellar populations of early type galaxies \citep{samiFP}.

\section*{Acknowledgements}
We are grateful to E. Emsellem and the anonymous referee for pointing us to relevant literature and for comments which improved the presentation of our results. 

This work was supported in part by NSF grant AST-1816330 to MB.
HDS acknowledges support from Centro Superior de Investigaciones Científicas PIE2018-50E099.  FN acknowledges support from the National Science Foundation Graduate Research Fellowship (NSF GRFP) under Grant No. DGE-1845298.  He and RKS thank the Munich Institute for Astro- and Particle Physics (MIAPP) which is funded by the Deutsche Forschungsgemeinschaft (DFG, German Research Foundation) under Germany's Excellence Strategy – EXC-2094 – 390783311, for its hospitality in 2019.

Funding for the Sloan Digital Sky Survey IV has been provided by the Alfred P. Sloan Foundation, the U.S. Department of Energy Office of Science, and the Participating Institutions. SDSS acknowledges support and resources from the Center for High-Performance Computing at the University of Utah. The SDSS web site is www.sdss.org.

SDSS is managed by the Astrophysical Research Consortium for the Participating Institutions of the SDSS Collaboration including the Brazilian Participation Group, the Carnegie Institution for Science, Carnegie Mellon University, the Chilean Participation Group, the French Participation Group, Harvard-Smithsonian Center for Astrophysics, Instituto de Astrof{\'i}sica de Canarias, The Johns Hopkins University, Kavli Institute for the Physics and Mathematics of the Universe (IPMU) / University of Tokyo, Lawrence Berkeley National Laboratory, Leibniz Institut f{\"u}r Astrophysik Potsdam (AIP), Max-Planck-Institut f{\"u}r Astronomie (MPIA Heidelberg), Max-Planck-Institut f{\"u}r Astrophysik (MPA Garching), Max-Planck-Institut f{\"u}r Extraterrestrische Physik (MPE), National Astronomical Observatories of China, New Mexico State University, New York University, University of Notre Dame, Observat{\'o}rio Nacional / MCTI, The Ohio State University, Pennsylvania State University, Shanghai Astronomical Observatory, United Kingdom Participation Group, Universidad Nacional Aut{\'o}noma de M{\'e}xico, University of Arizona, University of Colorado Boulder, University of Oxford, University of Portsmouth, University of Utah, University of Virginia, University of Washington, University of Wisconsin, Vanderbilt University, and Yale University.





\bibliographystyle{mnras}
\bibliography{biblio} 




\appendix
%


\renewcommand\thefigure{A\arabic{figure}}

\begin{figure*}
  \centering
  \includegraphics[width=0.8\linewidth]{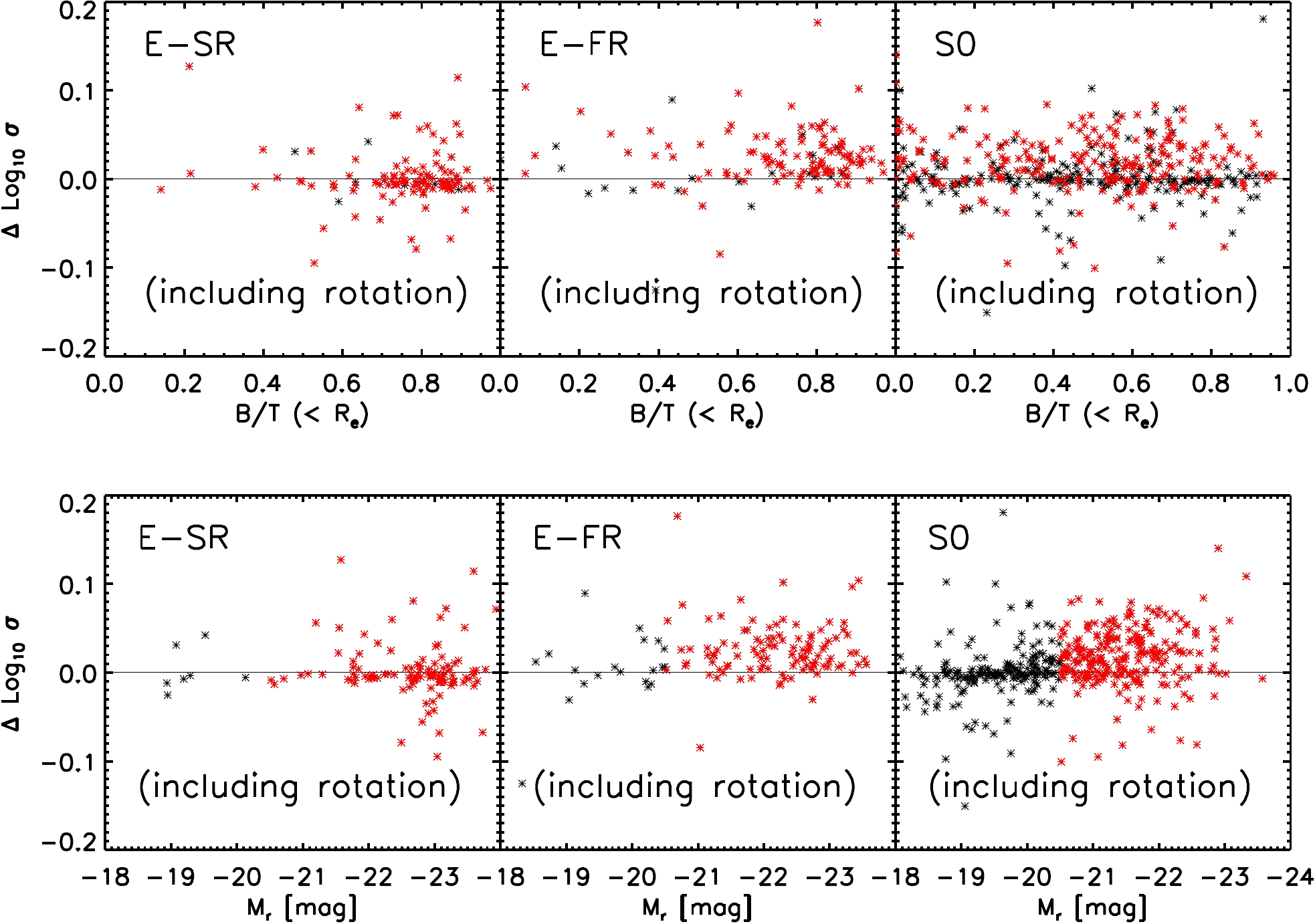}
  \caption{Ratio of $\sigma_e$ measured in the stacked spectrum to the square root of the (light-weighted) average of $V_{\rm rot}^2 + \sigma^2$ measured in the spaxels which contributed to the stack, shown as a function of B/T of the light within $R_e$ (top) and luminosity (bottom), for E-SRs, E-FRs and S0s (left to right).  Red symbols show the more luminous ($M_r<-21$) objects.}\label{fig:data}
  \vspace{0.5cm}
  \includegraphics[width=0.8\linewidth]{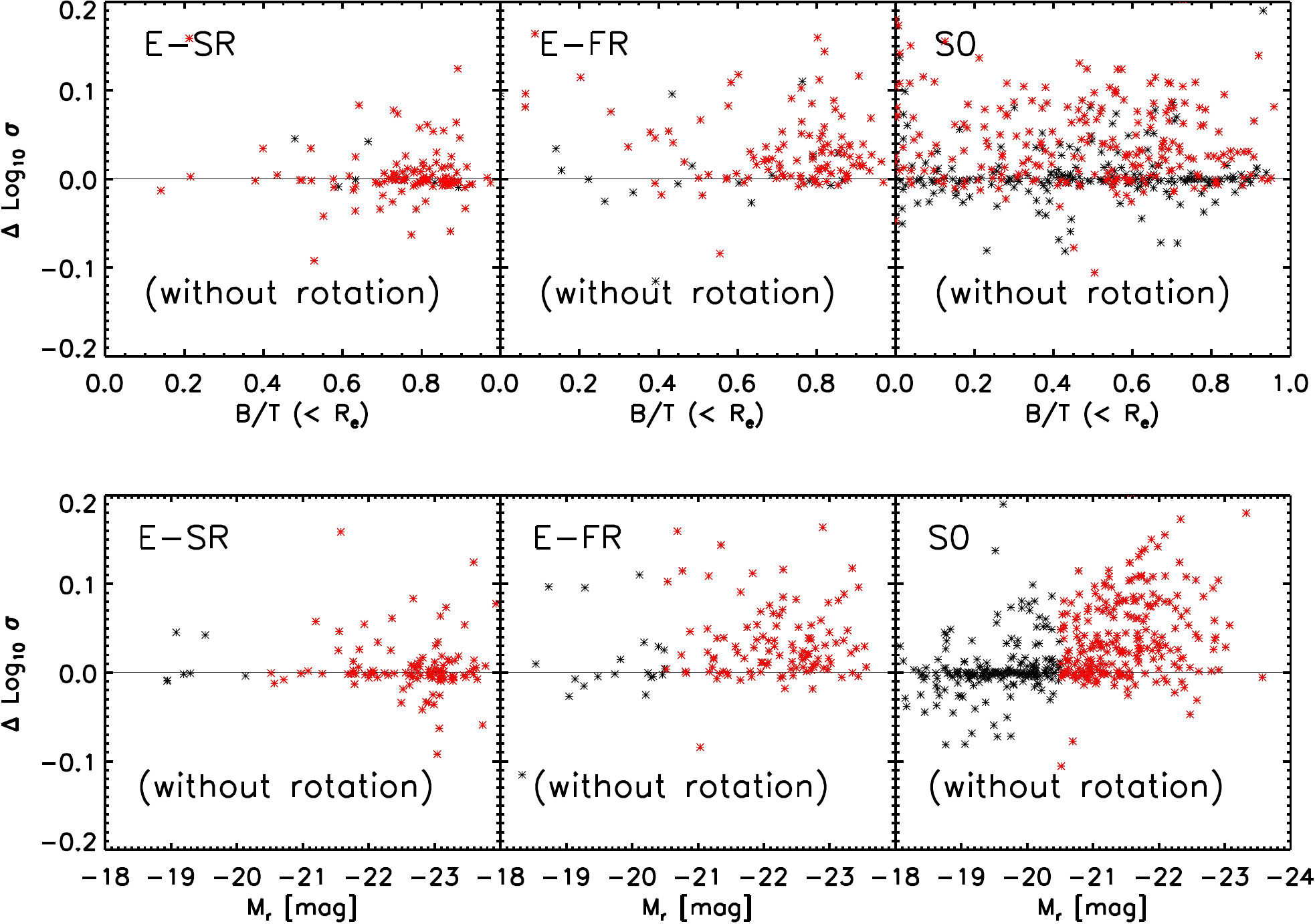}
  \caption{Same as previous Figure, but now for the ratio of $\sigma_e$ measured in rest-frame shifted and stacked spectra to the (light-weighted) average of $\sigma^2$ measured in the spaxels which contributed to the stack.}\label{fig:data-norot}
\end{figure*}

\section{Comparison of velocity dispersion estimates}\label{sec:sigVrot}

The main text described an estimate of the velocity dispersion which uses a stacked spectrum that will be somewhat contaminated by rotation.  In this Appendix we show how this estimate compares with the square root of the (light-weighted) average of $V_{\rm rot}^2 + \sigma^2$ measured in each spaxel which contributed to the stack.  Figure~\ref{fig:data} shows that the value measured from the stack tends to be slightly larger:  this difference is larger for S0s than for Es, and it increases slightly with luminosity.

To study if simply adding the rotation and dispersion in quadrature is the appropriate comparison, we have attempted to eliminate rotation from the discussion altogether as follows.  We first shift each spaxel to rest-frame -- using its reported $V_{\rm rot}$ value -- before stacking.  (Note that the ability to estimate a `no-rotation' $\sigma_e^2$ from these shifted and coadded spectra is unique to IFU-like datasets.)  We then estimate the dispersion of this stacked spectrum, and compare it to the (light-weighted) average of $\sigma^2$ of the spaxels which contributed to the stack.  Figure~\ref{fig:data-norot} shows that the difference between the two has increased rather than decreased.  

To understand why, we studied the two estimates in simulated spectra which we constructed as follows.  We selected a number of MaNGA galaxies randomly.  For each, we selected the spaxels within the ellipse of $R_{e,maj}$ (given the axis ratio $b/a$ of the galaxy), and computed the local value of B/T for each spaxel.  We modeled the bulge and disk components for each spaxel in the following way.  For simplicity, we assumed that the disk component rotates but has no dispersion, the bulge component does not rotate, and the two components have the same stellar populations.  Hence, to model the disc, we started from a stellar template spectrum, shifted it by the reported value of $V_{\rm rot}/(1 - B/T)$ and broadened it to have the SDSS instrumental dispersion of 70~km/s.  To model the bulge, we started from the same template spectrum and broadened it so that the light-weighted composite bulge+disk spectrum would have the reported dispersion.  We then added the two spectra to the stack, weighting them by the local value of the surface brightness.  In this way, our set of simulated spectra should have been made from reasonably realistic photometric and kinematic profiles.  (We note that they are not perfectly like the data:  it is unlikely that the bulge and disk components of real galaxies have similar stellar populations, and, while it is natural to hope that the bulge-disk decomposition in the photometry is reflected in the kinematics, we currently have no compelling evidence that this is indeed the case.)

We then treated the set of simulated spectra like the observed spectra.  The left-hand panel of Figure~\ref{fig:sim} shows the simulated analog of Figure~\ref{fig:data}.  It shows that the simulated spectra only show a bias when B/T$\sim 0.5$.  I.e., only in the worst case scenario -- where the photometry indicates that the galaxy is made of two comparable components -- does $\sigma_e^2$ from the stack differ from the light-weighted average of $V_{\rm rot}^2 + \sigma^2$.  The right-hand panel of Figure~\ref{fig:sim} shows the simulated analog of Figure~\ref{fig:data-norot}, in which the effects of rotation should have been removed.  Clearly, $\sigma_e^2$ from the stack is in good agreement with the light-weighted average of $\sigma^2$ from the spaxels, whereas there are significant differences between the two in the MaNGA data.  

\setcounter{figure}{2}

\begin{figure}
  \centering
  \includegraphics[width=0.82\linewidth]{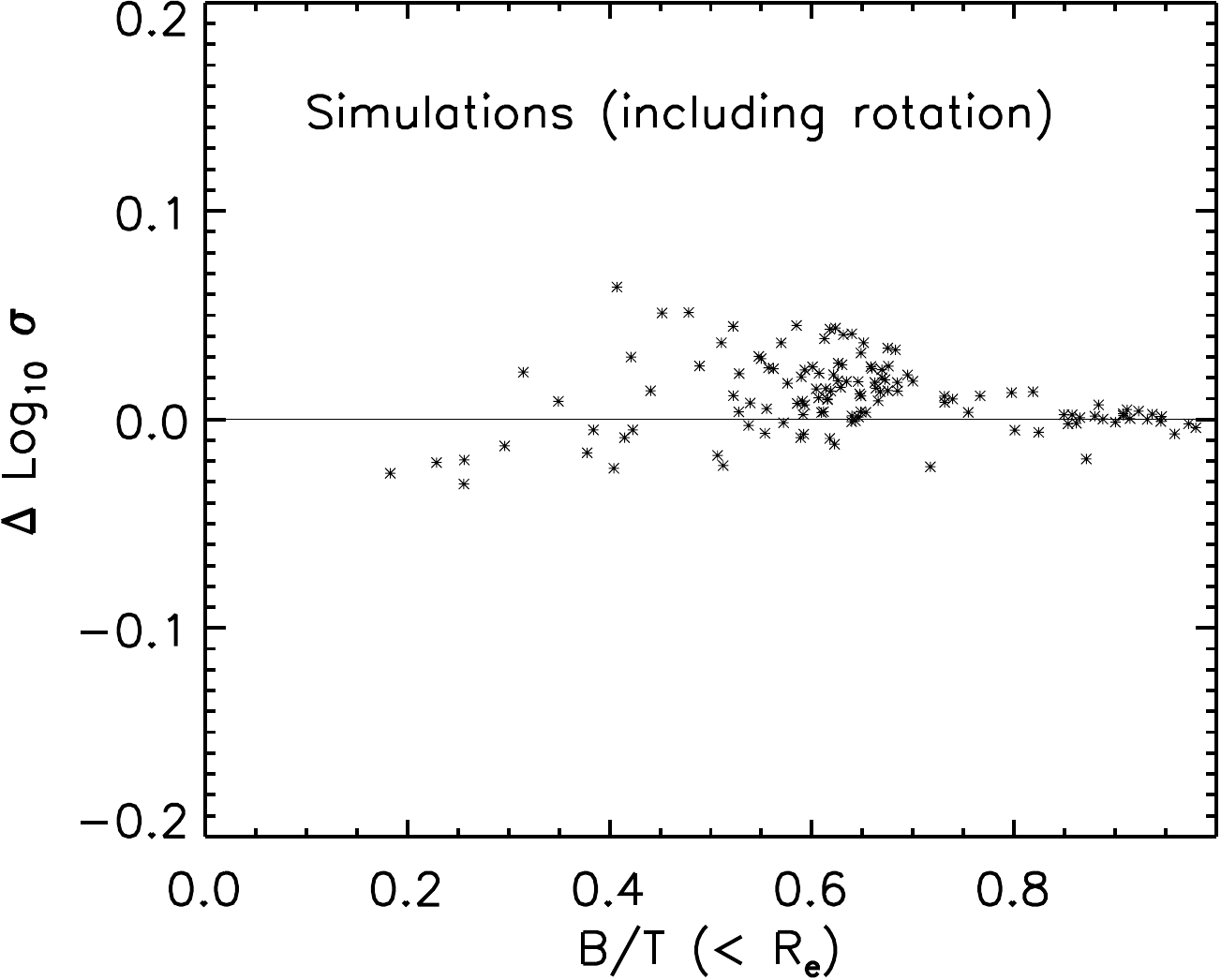}
  \includegraphics[width=0.82\linewidth]{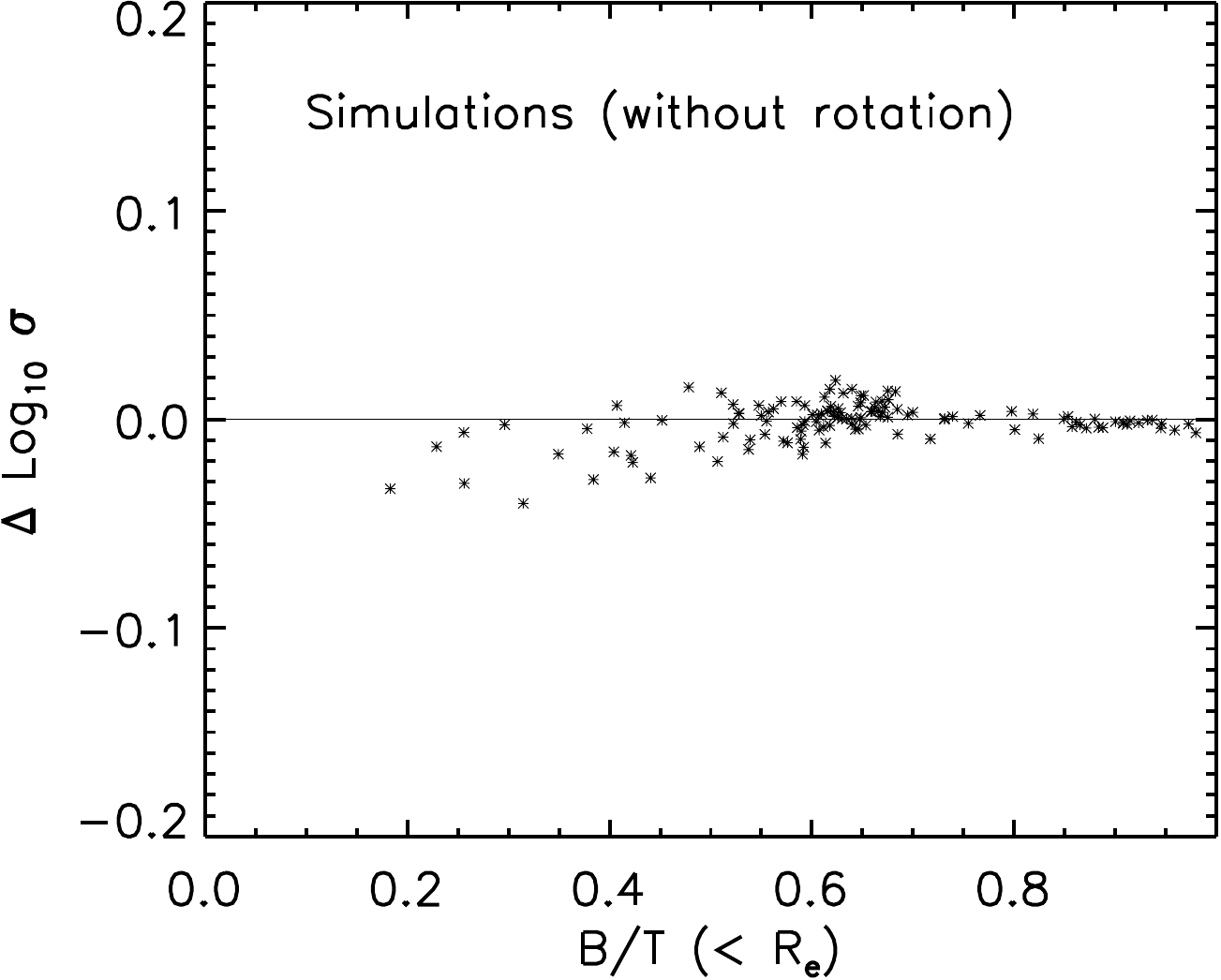}
  \caption{Top:  Ratio of $\sigma_e$ measured in a simulated stacked spectrum to the square root of the (light-weighted) average of $V_{\rm rot}^2 + \sigma^2$ measured in each spaxel which contributed to the stack, shown as a function of B/T of the light within $R_e$.  Bottom:  Same as top, but now showing ratio of $\sigma_e$ measured in the simulated rest-frame shifted and stacked spectrum to the light-weighted average of $\sigma$ measured in each spaxel which contributed to the stack.}
      \label{fig:sim}
\end{figure}

We believe this indicates that the measured values of $V_{\rm rot}$ (and hence $\sigma$) in each MaNGA spaxel are not accurate to the level we need.  This is because we can (and do) use the `correct' $V_{\rm rot}$ values to shift our simulated spectra back to restframe -- and when we do then the measurements in stacks agree with those from the spaxels.  However, in the data, there is no guarantee that the correct value has been used -- and the fact that there are relatively large differences between the stacks and the spaxels strongly suggests that the $V_{\rm rot}$ values are slightly biased.

One potential cause of bias is that the line shape associated with bulge+disk components is very non-Gaussian.  For this reason we have repeated the entire analysis allowing for non-Gaussian shapes (parametrized by third and fourth order Hermite-polynomials):  this does not resolve the discrepancy.

For this reason, the results in the main text use stacked spectra which have been broadened (slightly) by rotation.  Although not ideal -- they lead to small systematic trends with inclination (this is in addition to our discussion of the increase in scatter for S0s as $b/a\to 1$ -- right panel in the middle row of Figure~\ref{fig:FP} of the main text) -- they are at least closer to the measurements in the literature to date which do not require IFU data.  

\section{The FP with normalized variables}\label{sec:FPnorm}
It is an under appreciated fact that the detailed shape of the FP depends on the choice of units.  While it is conventional to work with size in kpc and velocities in km/s, there are two commonly used choices for the surface brightness:  $I_e$ in physical units or in magnitudes.  Since $\mu_e\propto -2.5\log_{10}I_e$, this choice only scales the inferred value of $\beta$ (by a factor of $-2.5$).  This is a trivial complication if one is only interested in describing one of the eigenvectors (typically the one normal to the plane).  However, this choice is crucial for relating the coefficients of this eigenvector to the two that lie in the plane.  

To make this point, let 
\begin{equation}
 \bmath{C_{\rm I}} = 
    \begin{pmatrix}
           C_{\rm VV} & C_{\rm VI} & C_{\rm VR}\\
           C_{\rm VI} & C_{\rm II} & C_{\rm IR}\\
           C_{\rm VR} & C_{\rm IR} & C_{\rm RR}\\
    \end{pmatrix}
 \label{eq:Ci}
\end{equation}
denote the covariance matrix of the three observables.  \cite{fp2012} provide explicit expressions for the eigenvalues and eigenvectors in terms of the $C_{ij}$.  In the present context it is particularly useful to write the two eigenvectors in the plane using the coefficients $\alpha$ and $\beta$ of the eigenvector normal to the plane and one extra coefficient $\kappa$:
\begin{equation}
  \label{eq:FPvectors}
 \bmath{\Lambda_{\rm I}} = 
    \begin{pmatrix}
       \kappa &  (1-\kappa\alpha)/\beta & 1 \\
       \kappa\alpha - \beta^2 - 1 & \beta(\kappa+\alpha) & \kappa\alpha^2 + \kappa\beta^2 -\alpha \\
      -\alpha &  -\beta & 1 
    \end{pmatrix}
    \begin{pmatrix}
         \bmath{v}\\
         \bmath{i}\\
         \bmath{r}
    \end{pmatrix},
\end{equation}
where $(\bmath{v},\bmath{i},\bmath{r})$ are unit vectors in the $(\log_{10}\sigma_e,\log_{10}I_e,\log_{10}R_e)$ directions.  The top row is the eigenvector with the largest eigenvalue; we refer to it as lying `along' the plane.  The middle row describes the eigenvector `across' the plane.  In general, all three eigenvectors have nontrivial dependence on $\bmath{v}$ and $\bmath{i}$.

An easy way to understand this strucure is to start with the fact that the bottom row gives the eigenvector with the smallest eigenvalue; it is normal to the plane, so it is orthogonal to the other two eigenvectors.  If we assume that the coefficient of $\bmath{v}$ for the vector along the FP is $\kappa$, then $\bmath{\Lambda}_1\cdot\bmath{\Lambda}_3 = 0$ sets the coefficient of $\bmath{i}$.  Similarly, $\bmath{\Lambda}_1\cdot\bmath{\Lambda}_2 = 0$ and $\bmath{\Lambda}_3\cdot\bmath{\Lambda}_2 = 0$ set the two coefficients of the vector across the FP.  In this way, the six independent numbers which define $\bmath{C_{\rm I}}$ are now encoded in the three eigenvalues and the three coefficients $\alpha,\beta$ and $\kappa$.  

In the SDSS 
\begin{equation}
 \bmath{C_{\rm I}} = 
    \begin{pmatrix}
           0.0205 & -0.0002 & 0.0255\\
          -0.0002 &  0.0816 & -0.0621\\
           0.0255 & -0.0621 & 0.0869\\
    \end{pmatrix}.
 \label{eq:Csdss}
\end{equation}
This matrix has $C_{\rm VI}\approx 0$, $C_{\rm II}\approx C_{\rm RR}$, and $C_{\rm IR}\approx -C_{\rm VR}$, with eigenvalues 0.149, 0.037 and 0.002 and eigenvectors given by equation~(\ref{eq:FPvectors}) with $(\alpha,\beta,\kappa) = (1.41,-0.78,0.2)$.

We will use $\bmath{C_\mu}$ to denote the corresponding matrix when $\log_{10} I_e$ is replaced by $\mu_e$.  I.e., $\bmath{C_\mu}$ is like $\bmath{C_{\rm I}}$ but with elements $C_{\mu\rm R} = -2.5\,C_{\rm IR}$, $C_{\rm V\mu} = -2.5\,C_{\rm VI}$, and $C_{\mu\mu} = 2.5^2\,C_{\rm II}$.  The eigenvalues of $\bmath{C_\mu}$ are $0.561$, $0.053$ and $0.002$.  If we use $\alpha_\mu$ and $\beta_\mu$ to denote the coefficients of the shortest eigenvector (in the basis $(\bmath{v},\bmath{u},\bmath{r})$, where $\bmath{u}$ is a unit vector in the $\mu_e$ direction), then $(\alpha_\mu,\beta_\mu,\kappa_\mu) = (1.41,0.31,0.05)$.  

Although the vector that is normal to the plane is similar to before -- $\beta_\mu=0.30 \approx -0.78/(-2.5)$ -- the eigenvector `along' the plane now has almost no coefficient in the $\bmath{v}$ direction.

The fact that $\kappa_\mu\approx 0$ suggested to \cite{efarVI} that there is some physical significance to separating the photometric parameters $R_e$ and $\mu_e$ from the spectroscopic one $\sigma_e$.  However, this is mainly a consequence of working with $\mu_e$ rather than $I_e$.  For instance, the axis ratios of $\bmath{C_\mu}$ are $10:1:0.04$; they were $4:1:0.05$ for $\bmath{C_{\rm I}}$, so the long axis is now much longer than the others.  This is simply because $C_{\mu\mu}$ is about $6\times$ larger than $C_{\rm RR}$.  E.g., had it been conventional to work with $5\log_{10}\sigma_e$ instead of $\log_{10}\sigma_e$, then $(\alpha,\beta,\kappa) = (0.25,-0.80,3.55)$ making the eigenvector along the plane $\Lambda_1$ have almost no component in the $I_e$ direction.  Thus, $\kappa=0$ results from the choice of units, and not from the underlying physics.

This raises the question of whether there is a more fundamental way to describe the FP and its eigenvectors.  The most natural choice is to normalize all quantities by their rms values.  The normalized $\bmath{C_{\rm I}}$ has ones along the diagonal and $(r_{\rm IR},r_{\rm VR},r_{\rm VI})\approx (-0.74,0.0.60,-0.01)$, where $r_{ij} = C_{ij}/\sqrt{C_{ii}C_{jj}}$.  
As a result, for the SDSS Es the eigenvalues are 1.95:1:0.05 (the sum of the eigenvalues must equal 3) and 
\begin{align}
  \bmath{\Lambda}_1 &\approx 1.1\bmath{\hat{i}} - \bmath{\hat{v}}/1.1 - 1.4{\bmath{\hat{r}}} \nonumber\\
  \bmath{\Lambda}_2 &\approx \bmath{\hat{i}}/1.1 + 1.1\bmath{\hat{v}} \\
  \bmath{\Lambda}_3 &\approx 1.1\bmath{\hat{i}} - \bmath{\hat{v}}/1.1 + 1.4\bmath{\hat{r}},\nonumber
\end{align}
where $\bmath{\hat{r}}=\bmath{r}/\sqrt{C_{\rm RR}}$ etc.
This is quite similar to equations~(53-55) of \cite{fp2012} which have $1.1\to 1$.  Evidently, in these normalized units, one eigenvalue is much smaller than the other two (as expected for a Plane) -- and both (approximately) have $\bmath{\hat{i}} + \bmath{\hat{v}}$ as the vector `across' the plane.  The structure of this normalized plane is a direct consequence of the fact that $I_e$ and $\sigma_e$ are almost uncorrelated ($r_{\rm IV}\approx 0$), whereas the other two correlations are nearly equal and opposite:  $r_{\rm IR}\approx -r_{\rm RV}$.  For the general case, when $r_{\rm IV}\approx 0$ but $r_{\rm IR}\ne -r_{\rm RV}$, see equations~(50-52) of \cite{fp2012}.

For completeness, we note that, for the MaNGA Es (E-SRs + E-FRs), the correlation coefficients are $(-0.75,0.0.70,-0.10)$, and the eigenvectors of the normalized FP are well approximated by
\begin{align}
  \bmath{\Lambda}_1 &\approx 1.1\bmath{\hat{i}} - \bmath{\hat{v}} - 1.4\bmath{\hat{r}}\nonumber\\
  \bmath{\Lambda}_2 &\approx \bmath{\hat{i}}/1.1 + \bmath{\hat{v}}\\
  \bmath{\Lambda}_3 &\approx \bmath{\hat{i}} - \bmath{\hat{v}}/1.1 + 1.4\bmath{\hat{r}} ,\nonumber
\end{align}
with eigenvalues 2.08:0.9:0.02.  Again, the fact that the vector across the plane is approximately $\bmath{\hat{i}} + \bmath{\hat{v}}$ indicates that $r_{\rm VI}\approx 0$, whereas $r_{\rm IR}\approx -r_{\rm VR}$.  As \cite{fp2012} have emphasized, this is the `units-independent' physical information content of the structure of the Plane which galaxy formation models should strive to reproduce.

\bsp	
\label{lastpage}
\end{document}